\documentclass[onecolumn, trackchanges]{aastex631}
\usepackage{enumerate}
\usepackage{import}
\usepackage{xspace}
\usepackage{color}
\usepackage{amsmath}
\usepackage{appendix}
\usepackage{hyperref}

\newcommand{\rjup}{\ensuremath{R_J}}
\newcommand{\ms}{\ensuremath{\mathrm{m\,s^{-1}}}}
\newcommand{\cms}{\ensuremath{\mathrm{cm\,s^{-1}}}}

\shortauthors{Blunt et al.}
\shorttitle{V1298~Tau and Overfitting}

\begin{document}

\pagenumbering{arabic}

\title{Overfitting Affects the Reliability of Radial Velocity Mass Estimates of the V1298~Tau Planets}

\author[0000-0002-3199-2888]{Sarah Blunt}
\affiliation{Department of Astronomy, California Institute of Technology, Pasadena, CA 91125, USA}

\author[0000-0002-9540-853X]{Adolfo Carvalho}
\affiliation{Department of Astronomy, California Institute of Technology, Pasadena, CA 91125, USA}

\author[0000-0001-6534-6246]{Trevor J. David}
\affiliation{Center for Computational Astrophysics, Flatiron Institute, New York, NY 10010, USA}

\author[0000-0002-5627-5471]{Charles Beichman}
\affiliation{NASA Jet Propulsion Laboratory: Pasadena, CA, US}

\author[0000-0003-1848-2063]{Jon K. Zink}
\affiliation{Department of Astronomy, California Institute of Technology, Pasadena, CA 91125, USA}
\altaffiliation{NHFP Sagan Fellow}

\author[0000-0002-5258-6846]{Eric Gaidos}
\affiliation{Department of Earth Sciences, University of Hawai'i at Mānoa, Honolulu, HI 96822 USA}

\author[0000-0003-0012-9093]{Aida Behmard}
\affiliation{Division of Geological and Planetary Sciences,
1200 E California Blvd, Pasadena, CA, 91125, USA}

\author[0000-0002-0514-5538]{Luke G. Bouma}
\altaffiliation{51 Pegasi b Fellow}
\affiliation{Department of Astronomy, California Institute of Technology, Pasadena, CA 91125, USA}

\author[0000-0002-7713-5937]{Devin Cody} 
\affiliation{Google LLC, 1600 Amphitheater Parkway, Mountain View, CA 94043, USA}

\author[0000-0002-8958-0683]{Fei Dai} 
\affiliation{Division of Geological and Planetary Sciences,
1200 E California Blvd, Pasadena, CA, 91125, USA}
\affiliation{Department of Astronomy, California Institute of Technology, Pasadena, CA 91125, USA}

\author[0000-0002-9328-5652]{Daniel Foreman-Mackey}
\affiliation{Center for Computational Astrophysics, Flatiron Institute, New York, NY 10010, USA}

\author[0000-0003-4976-9980]{Sam Grunblatt}
\affiliation{American Museum of Natural History, 200 Central Park West, Manhattan, NY 10024, USA}
\affiliation{Center for Computational Astrophysics, Flatiron Institute, New York, NY 10010, USA}
\affiliation{Department of Physics and Astronomy, Johns Hopkins University, 3400 N Charles St, Baltimore, MD 21218, USA}
\altaffiliation{Kalbfleisch Fellow}

\author[0000-0001-8638-0320]{Andrew W.\ Howard}
\affiliation{Department of Astronomy, California Institute of Technology, Pasadena, CA 91125, USA}

\author{Molly Kosiarek}
\altaffiliation{NSF Graduate Research Fellow}
\affiliation{Department of Astronomy and Astrophysics, University of California, Santa Cruz, CA 95064, USA}

\author[0000-0002-5375-4725]{Heather A. Knutson} 
\affiliation{Division of Geological and Planetary Sciences,
1200 E California Blvd, Pasadena, CA, 91125, USA}

\author[0000-0003-3856-3143]{Ryan A. Rubenzahl}
\altaffiliation{NSF Graduate Research Fellow}
\affiliation{Department of Astronomy, California Institute of Technology, Pasadena, CA 91125, USA}


\author[0000-0001-7708-2364]{Corey Beard}
\affiliation{Department of Physics \& Astronomy, The University of California, Irvine, Irvine, CA 92697, USA}

\author[0000-0003-1125-2564]{Ashley Chontos}
\altaffiliation{Henry Norris Russell Fellow}
\affiliation{Department of Astrophysical Sciences, Princeton University, 4 Ivy Lane, Princeton, NJ 08540, USA}

\author[0000-0002-8965-3969]{Steven Giacalone}
\affil{Department of Astronomy, University of California Berkeley, Berkeley, CA 94720, USA}

\author[0000-0003-3618-7535]{Teruyuki Hirano}
\affiliation{Astrobiology Center, 2-21-1 Osawa, Mitaka, Tokyo 181-8588, Japan}
\affiliation{National Astronomical Observatory of Japan, 2-21-1 Osawa, Mitaka, Tokyo 181-8588, Japan}

\author[0000-0002-5099-8185]{Marshall C. Johnson} 
\affiliation{Department of Astronomy, The Ohio State University, 4055 McPherson Laboratory, 140 West 18$^{\mathrm{th}}$ Ave., Columbus, OH 43210 USA}

\author[0000-0001-8342-7736]{Jack Lubin}
\affiliation{Department of Physics \& Astronomy, The University of California, Irvine, Irvine, CA 92697, USA}

\author[0000-0001-8898-8284]{Joseph M. Akana Murphy}
\altaffiliation{NSF Graduate Research Fellow}
\affiliation{Department of Astronomy and Astrophysics, University of California, Santa Cruz, CA 95064, USA}

\author[0000-0003-0967-2893]{Erik A Petigura}
\affiliation{Department of Physics \& Astronomy, University of California Los Angeles, Los Angeles, CA 90095, USA}

\author[0000-0002-4290-6826]{Judah Van Zandt}
\affiliation{Department of Physics \& Astronomy, University of California Los Angeles, Los Angeles, CA 90095, USA}

\author[0000-0002-3725-3058]{Lauren Weiss}
\affil{Department of Physics and Astronomy, University of Notre Dame, Notre Dame, IN 46556, USA}




\begin{abstract}

Mass, radius, and age measurements of young ($\lesssim$ 100 Myr) planets have the power to shape our understanding of planet formation. However, young stars tend to be extremely variable in both photometry and radial velocity, which makes constraining these properties challenging. The V1298~Tau system of four $\sim$0.5~R$_{\rm J}$ planets transiting a pre-main sequence star presents an important, if stress-inducing, opportunity to directly observe and measure the properties of infant planets. \citet[hereafter SM21]{SM21} published radial-velocity-derived masses for two of the V1298~Tau planets using a state-of-the-art Gaussian Process regression framework. The planetary densities computed from these masses were surprisingly high, implying extremely rapid contraction after formation in tension with most existing planet formation theories. In an effort to further constrain the masses of the V1298~Tau planets, we obtained 36 RVs using Keck/HIRES, and analyzed them in concert with published RVs and photometry. Through performing a suite of cross validation tests, we found evidence that the preferred model of SM21 suffers from overfitting, defined as the inability to predict unseen data, rendering the masses unreliable. We detail several potential causes of this overfitting, many of which may be important for other RV analyses of other active stars, and recommend that additional time and resources be allocated to understanding and mitigating activity in active young stars such as V1298~Tau. 

\end{abstract}

\section{Introduction}

\subsection{Young Planets as Probes of Formation}

Planet formation is an uncertain process. Giant planets are thought to form with large radii, inflated due to trapped heat, then cool and contract over the first few hundred Myr of their lives \citep{Marley:2007aa}. However, the accretion efficiency of the formation process, which sets the planets' initial entropy and radii, spans orders of magnitude of uncertainty. The processes sculpting the post-formation masses and radii of smaller terrestrial exoplanets are also uncertain. Young, terrestrial planets also have uncertain initial entropies, and for highly irradiated planets, the unknown rate of photoevaporation (itself due to uncertainties in  a planet's migration history, among other physical unknowns) during and after formation compounds this ambiguity (\citealt{Lopez:2012aa}, \citealt{Owen:2013aa}, \citealt{Chen:2016aa}, \citealt{Owen:2020aa}).

Measuring the masses, radii, and ages of newly-formed planets presents a path forward \citep{Owen:2020aa}. Young moving groups provide rigorous age constraints, and relatively model-independent methods of measuring planetary radii exist for both young directly-imaged and transiting planets (for transiting planets in particular, only stellar radius model dependencies impact the inferred planetary radius). However, in both situations, few model-independent mass measurements exist. For transiting planets, there are two complementary methods for measuring planetary masses: transit timing variations (TTVs), and  stellar radial velocity (RV) timeseries. 

Measuring RV masses of young planets is a difficult task, so some advocate to rely on transit timing variations (TTVs) alone to measure masses of young planets. However, not all planets transit, and only planets in multi-planet systems at or near mean motion resonance exhibit TTVs \citep{Fabrycky:2014aa}. Even in systems that do, individual TTV mass posteriors are often covariant, since TTVs to first order constrain the planetary mass ratio (\citealt{Lithwick:2012aa}, \citealt{Petigura:2020aa}). In an ideal scenario, both RVs and TTVs would be used to jointly constrain planetary masses in a given system, reducing posterior uncertainty and TTV degeneracies.

\subsection{Stellar Activity \& Overfitting}

As the instrumental errors of extremely precise RV instruments approach 10 cm s$^{-1}$, and as the RV community begins to target more active stars, accurately modeling astrophysical noise is becoming more and more critical. Young stars present a particular challenge. These are highly magnetically active \citep{Johns-Krull_BFieldsTTauStars_2007ApJ}, with starspots that occupy significant fractions of the stellar surface and induce RV variations on the order of $\sim$km s$^{-1}$ \citep{Saar_RVVariation_1997ApJ}. These RV variations are hundreds of times larger than the activity signals of older quiet stars typically targeted by RV surveys and complicate the detection of planet-induced Doppler shifts from even close-in Jupiter-mass planets \citep[e.g.,][]{huerta_starspot-induced_2008, Prato_VisIR_2008ApJ}. 

Other assumptions and/or information can be leveraged to model the activity signal, even if the signal isn't easily understandable from the RVs themselves. A widely used practice involves independently constraining the rotation period from a photometric timeseries, then using an informed prior on the rotation period to model the RVs (e.g., \citealt{Grunblatt:2015aa}). Other related examples include specifying a quasi-periodic kernel for a Gaussian Process regression (GPR) model (i.e., assuming that the stellar activity has a quasi-periodic form), or modeling the RVs jointly with other datasets. The latter approach achieves better model constraints either by explicitly modeling the relationship between the datasets (e.g., \citealt{Rajpaul:2015aa}) or by sharing hyperparameters between datasets (e.g., \citealt{Grunblatt:2015aa}, \citealt{Morales:2016aa}).  

As is true for every model-fitting process, misspecifying the stellar activity model (i.e., fitting a model that is not representative of the process that generated the data) or allowing too many effective degrees of freedom can lead to overfitting.

Overfitting is a concept ubiquitous in machine learning, and in particular is often used to determine when a model has been optimally trained. One algorithm for determining whether a model is overfitting is as follows\footnote{See also \citet{Cale:2021aa}, who define overfitting in terms of reduced $\chi^2$ in the context of RV activity modeling}: divide the data into a ``training'' set and an ``evaluation'' set (a common split is 80\%/20\%), and begin optimizing the model using just the training set. At each optimization step, calculate the goodness-of-fit metric for the model on the evaluation set, which is otherwise omitted from the training process altogether. This method of evaluating a model’s ability to successfully predict new, or ``out-of-sample,'' data is known as cross validation (CV). 

The classic observed behavior is that the goodness-of-fit metrics for both the training and evaluation set improve as the model fits the training data better and better. At a certain point, the model begins to overfit to the training data, and the goodness-of-fit metric for the evaluation data worsens. This is because the model parameters have begun to reproduce the noise in the training set, at the expense of reproducing the signal common to both datasets. A model that is overfitting, then, can be defined as one that predicts the observations in a training set better than those in an evaluation set. An overfitting model fits aspects of the data that are not predictable or common to the entire data set, e.g., noise.

The optimally trained model is selected not by its performance relative to the \textit{training} data, but by its performance relative to the \textit{evaluation} data, which was omitted from the training process altogether. Making an analogy to Bayesian model comparison, we could imagine a similar process where the goodness-of-fit is evaluated for an ``evaluation set'' left out of the training process (i.e., posterior computation using MCMC, nested sampling, etc.) for a series of models. One benefit of this method over, e.g., formal Bayesian model comparison is that it also provides an easily-interpretable absolute metric for how well the model fits the data: if the evaluation set goodness-of-fit is significantly worse than that of the training set, we know the model is misspecified, even if it has (comparatively) the lowest Bayesian evidence.

In this study, we apply the CV technique as defined above to evaluate the predictiveness of one particular model fit to one particular star. This is intended as a case study, aiming to inspire further investigation into the extent of and causes of overfitting in RV modeling of young, active stars.

\subsection{V1298 Tau}

V1298 Tauri (hereafter V1298 Tau) is a young system of four $\gtrsim0.5$ \rjup{} planets transiting a K-type pre-main sequence (PMS) star (\citealt{David2019aa}, \citealt{David2019ab}). Very few transiting planets have been discovered around PMS stars (other notable systems being AU Mic, \citealt{Plavchan2020aa}, \citealt{Cale:2021aa}, \citealt{Zicher:2022aa}, \citealt{Klein:2022aa}; K2-33, \citealt{David2016aa}; DS Tuc, \citealt{Newton:2019aa}, \citealt{Benatti:2019aa}; HIP 67522 \citealt{Rizzuto:2020aa}; Kepler 1627A, \citealt{Bouma:2022aa}; and TOI 1227, \citealt{Mann:2022aa}). \citet{David2019aa} reported the discovery of V1298 Tau b, a 0.9 \rjup{} object with an orbital period of 24d. \citet{David2019ab} discovered three additional planets in the system: V1298 Tau c at 8d, V1298 Tau d at 12d, and a single-transiting object, V1298 Tau e. 

It is unclear from their radii ($0.5-1\rm{R}_J$) alone whether these planets are gas giants that contracted rapidly after forming, or young terrestrial or mini-Neptune planets, which will lose a large fraction of their atmospheres to photoevaporation \citep{Owen:2013aa} and/or core-powered mass loss \citep{Ginzburg2018aa} as they age.

\citet{SM21}, hereafter SM21, presented over 100 RVs of the system with four different instruments, and used a suite of Gaussian Process (GP) models in an effort to isolate the RV signals of the outer two transiting planets (b and e), finding masses of 0.64 and 1.16 $M_J$ for each respectively. Combined with radii of 0.91 and 0.78 $R_J$, this implied high densities of 1.2 and 3.6 g cm$^{-3}$. Such high densities would require rapid contraction after within the 23 Myr lifetime of the system and place the outer V1298 Tau planets at the upper density boundary of even mature field exoplanets, where few theories predict planets to exist. Since V1298 Tau e only transited once while observed with K2, SM21 did not have a period constraint from transits, and derived a period of 40.2~$\pm~0.9$~d  from their RV measurements. After the publication of SM21, \citet{Feinstein2022aa} published updated ephemerides of all four planets using TESS photometry, including a second transit of planet e. They placed a strict lower limit of 42~d on V1298 Tau e's period, in tension with the value from SM21. In addition, \citet{Tejada:2022aa} performed a dynamical stability analysis using the mass measurements reported in SM21, finding that 97\% of system configurations consistent with the SM21 posteriors are gravitationally unstable over the lifetime of the system. 

SM21 performed a rigorous and state-of-the-art analysis, comparing several complex models with Bayesian methods. However, several independent lines of evidence appear to call the masses they report into question: the tension with formation theory discussed in SM21, the updated planet e orbital period of \citet{Feinstein2022aa}, and the improbability of long-term stability derived by \citet{Tejada:2022aa}. 

\subsection{This Paper}

The purpose of this paper is two-fold: 1) to demonstrate the use of CV to show that the preferred model of SM21 is overfitting the RVs, and 2) to point out several potential causes of the overfitting, which are not unique to SM21 but common throughout the literature. We do not attempt to update the mass estimates for the V1298 Tau planets in this paper. We also do not attempt to prove that the mass estimates published in SM21 are incorrect, but instead seek to call into question their reliability. Our argument is that future joint models of the stellar activity and planetary signals of V1298~Tau will need to prove their predictiveness in order to be trustworthy. 

The structure of this paper is as follows: in Section \ref{sec:data}, we review the literature data scrutinized in this paper and describe one additional contemporaneous RV dataset taken with Keck/HIRES. In section \ref{sec:overfitting}, we demonstrate that the preferred model of SM21 is overfitting. Section \ref{sec:why} discusses several potential causes of this overfitting, and advises modelers on how to detect and/or avoid these subtle pitfalls. In particular, Section \ref{sec:diffrot} argues that differential rotation is an important effect for V1298 Tau, and must be modeled carefully. We conclude in Section \ref{sec:discuss}. We also provide an appendix that provides a geometric interpretation of how GPR penalizes complexity. All of the code to create the plots in this work are publicly available on GitHub\footnote{\href{https://github.com/sblunt/V1298Tauri}{https://github.com/sblunt/V1298Tauri}}.

\section{Data}
\label{sec:data}

\begin{figure*}
    \centering
    \hspace*{-0.5in}
    \includegraphics{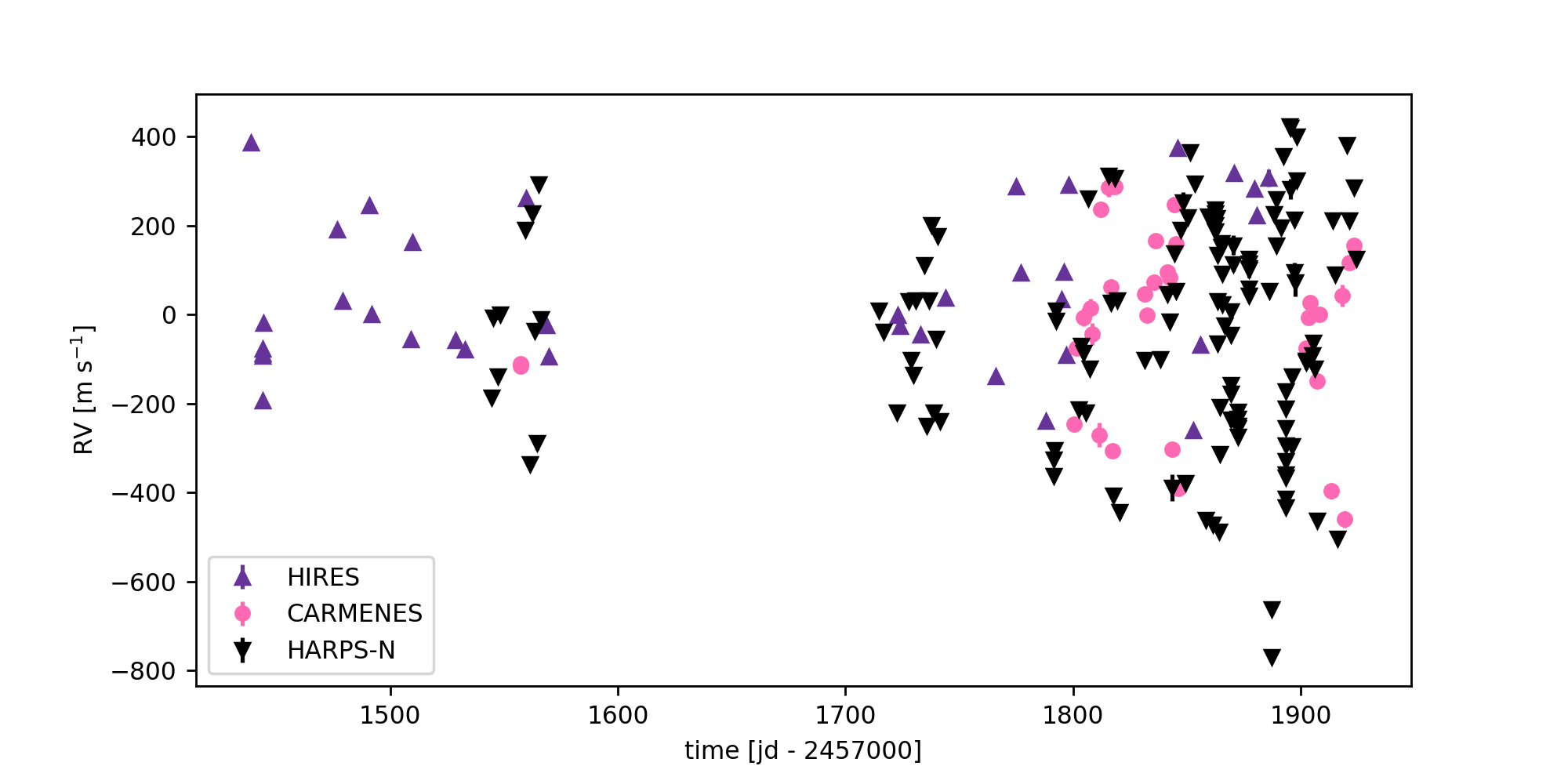}
    \caption{A tour of the RVs scrutinized in this study. The CARMENES and HARPS-N RVs are published in SM21, and the HIRES RVs are new in this study. Takeaway: the RV variability of V1298~Tau is hundreds of m/s, which is similar across all three instruments. The variability is significantly greater than the instrumental errors (which are included, but too small to see for the majority of points on this plot).}
    \label{fig:rvs}
\end{figure*}

Throughout this paper, we reference several data sets: three photometric time series measured by different instruments and three RV timeseries derived from spectra measured by different instruments. Each dataset is detailed in the subsections below. All of the photometry is shown in Figure \ref{fig:phot}, and all of the RVs are shown in Figure \ref{fig:rvs}. 

\subsection{K2 photometry}

We downloaded EVEREST-processed (\citealt{Luger:2016aa}, \citealt{Luger:2018aa}) K2 lightcurves for V1298~Tau using the \texttt{lightkurve} package \citep{lightkurve:2018aa}. We used built-in \texttt{lightkurve} functions to remove nans, remove outliers, and normalize the data.

\subsection{LCO photometry}

We obtained ground-based LCO photometry verbatim from SM21. 

\subsection{TESS photometry}

We obtained TESS lightcurves from \citet{Feinstein2022aa}, who combined timeseries photometry of V1298~Tau from TESS Sectors 43 and 44. \citet{Feinstein2022aa} used the 2-minute light curve created by the Science Processing Operations Center pipeline (SPOC; \citealt{Jenkins:2016aa}), and binned those observations to 10 mins. We normalized the data for each TESS orbit separately, following \citet{Feinstein2022aa}. 

\subsection{SM21 RVs}

We obtained CARMENES and HARPS-N RVs directly from SM21. We note that SM21 excluded infrared-arm CARMENES RVs in its analysis, and we do the same here. 

HARPS-N RVs are wavelength calibrated using a ThAr lamp, and the HARPS-N spectrograph covers 360-690 nm.

The visible arm of the CARMENES instrument covers the spectral range 520-960 nm, and spectra from this instrument are wavelength calibrated using a Fabry-Perot etalon, anchored using hollow cathode lamps.

\subsection{Keck/HIRES RVs}

Between November 16, 2018, and February 6, 2020, we obtained 36 RVs using the HIRES spectrograph on the Keck I telescope \citep{Vogt:1994aa}. Wavelength calibration was performed by passing starlight through a warm iodine cell, and data reduction was performed using the California Planet Search pipeline described in \citet{Howard:2010aa}, which is adapted from \citet{Butler:1996aa}. All HIRES RVs used in this study are given in Table \ref{table:rvs}. Some of these RVs were previously published in \cite{Johnson:2022aa}, and the processing is identical in that paper and this. The same stellar template, constructed from two stellar spectra taken on 24 Oct 2019 UT without the iodine in the light path, was used to derive RVs in both studies. In-transit RVs from that study have been excluded here. Spectra were typically taken using the C2 decker (14'' x 0.861''), which enables sky subtraction, and is the CPS HIRES observer ``decker of choice'' for stars fainter than V$\sim$10. However, a CPS HIRES observer ``rule of thumb'' is to use the shorter B5 decker (3.5'' x 0.861'') in poor seeing conditions, as the Doppler pipeline sky subtraction algorithm is unreliable when the stellar PSF fills the slit. Sky subtraction is not performed under such conditions. Accordingly, 7 RVs published here were calculated from spectra using the B5 decker. In both modes, HIRES has a resolving power of $\sim$60,000, and the iodine cell spectral grasp translates to contributions to the RV from wavelengths between 500 and 620 nm \citep{Butler:1996aa}. 

\begin{deluxetable*}{ccc}
\tablecaption{New HIRES RV Data. Epochs are reported in units of BJD\_TDB, with a 2457000 offset applied. \label{table:rvs}}
\tablehead{\colhead{jd - 2457000} & \colhead{RV [m s$^{-1}$]} & \colhead{RV error [m s$^{-1}$]}
}
    \startdata
1438.9463 & 387.47 & 7.88 \\
1443.8205 & -193.32 & 8.53 \\
1443.9573 & -87.96 & 7.94 \\
1443.9682 & -77.17 & 7.07 \\
1443.9792 & -92.67 & 8.61 \\
1444.1566 & -19.40 & 8.40 \\
1476.8089 & 190.28 & 11.71 \\
1479.0060 & 30.83 & 10.27 \\
1490.7912 & 246.54 & 10.05 \\
1491.7549 & 0.31 & 8.80 \\
1508.9196 & -56.33 & 9.53 \\
1509.7900 & 163.24 & 12.03 \\
1528.7841 & -58.17 & 9.75 \\
1532.7701 & -78.17 & 10.47 \\
1559.7316 & 261.44 & 9.53 \\
1568.7290 & -23.72 & 9.68 \\
1569.7376 & -94.99 & 11.02 \\
1723.1354 & -0.31 & 8.62 \\
1724.0917 & -26.27 & 9.65 \\
1733.0739 & -44.28 & 9.66 \\
1743.9965 & 37.42 & 10.46 \\
1765.9266 & -138.49 & 11.11 \\
1774.9154 & 287.77 & 9.43 \\
1777.0163 & 94.72 & 9.40 \\
1787.9817 & -239.82 & 10.01 \\
1794.8691 & 34.79 & 8.83 \\
1795.9062 & 95.70 & 8.25 \\
1796.9165 & -90.09 & 8.83 \\
1797.9532 & 292.26 & 8.13 \\
1845.8966 & 375.28 & 9.47 \\
1852.7705 & -259.55 & 9.59 \\
1855.8832 & -68.08 & 16.10 \\
1870.8459 & 318.54 & 8.90 \\
1879.8314 & 282.03 & 9.12 \\
1880.8314 & 222.22 & 8.49 \\
1885.8611 & 306.95 & 19.62 \\
\enddata
\end{deluxetable*}

\section{Cross Validation Tests}
\label{sec:overfitting}

Our intention in collecting additional RVs of V1298~Tau with HIRES was to jointly analyze these data together with literature data and update the masses published in SM21. However, early on in the analysis, we noticed clues that made us question our assumptions. In particular, the new data we had collected did not seem consistent with the models of SM21. In addition, many tested models converged on results that were physically unreasonable or clearly inconsistent with subsets of the data. We ultimately decided to test the predictive capability of the SM21 model that we were using as our starting point, as a check on our own assumptions. This section details the outcome of those experiments.

The main finding of this paper is that the median parameter estimate of the preferred model of SM21 (their \textbf{4p}$_{PQP2}$) is overfitting. For convenience, we will refer to this model throughout the rest of this paper as the ``SM21 preferred model.'' Showing that a point estimate is overfitting does not necessarily indicate that every model spanned by the posterior is overfitting. However, since the preferred model presented by SM21 (their figure 11) appears approximately Gaussian around the MAP estimates of the parameters relevant for us, (except for the kernel parameter C and the white noise jitter for CARMENES, which both peak at effectively 0), we assume that the MAP and median for this fit are close enough to make no difference, and that inferences made about the median fit hold true for other high-probability areas of parameter space. 

\begin{figure*}
    \centering
    \hspace*{-0.5in}
    \includegraphics{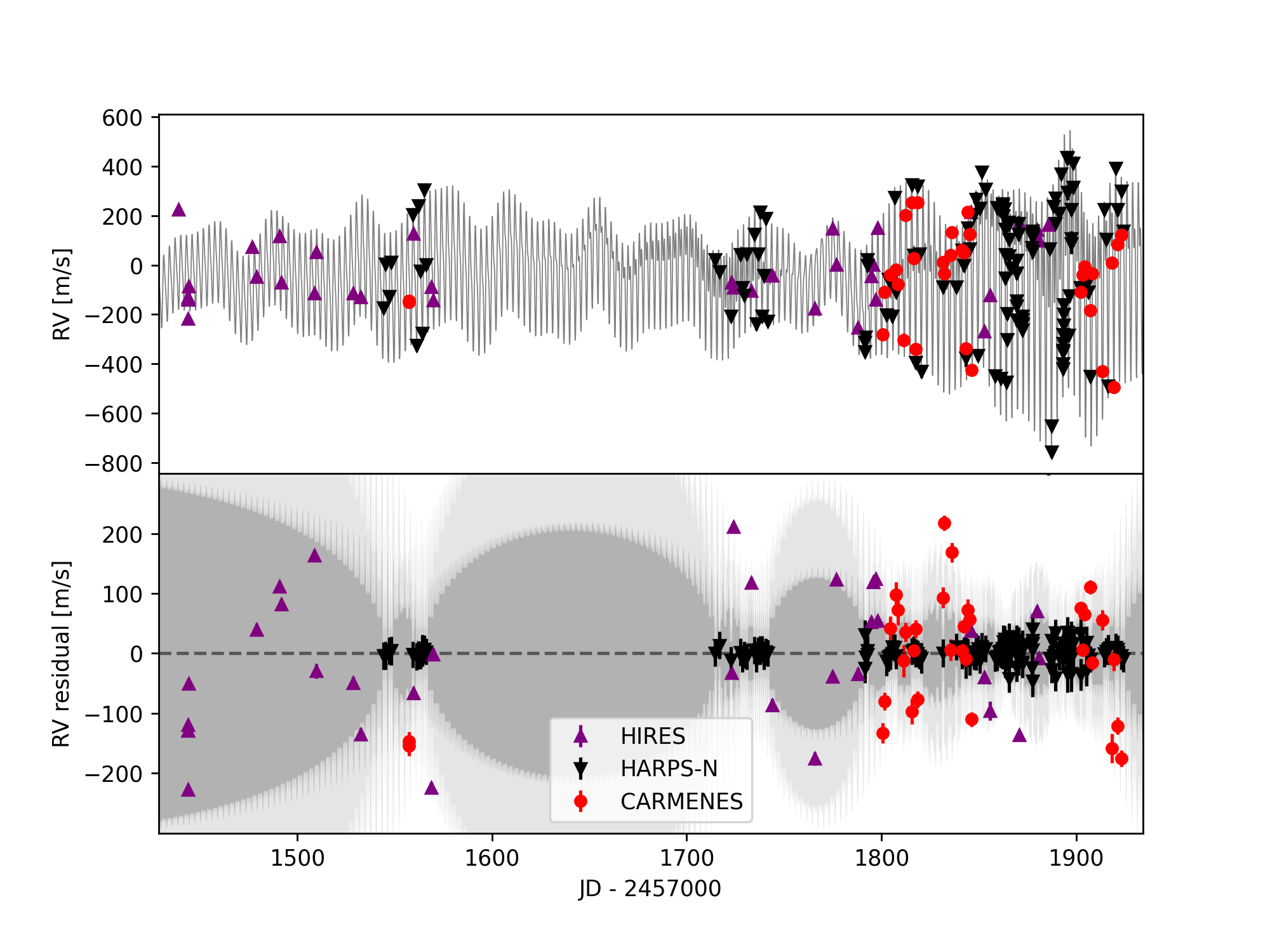}
    \caption{SM21 preferred model prediction and contemporaneous observed data. The HIRES data have been scaled and offset by linear parameters that minimize the residual spread with respect to the GP model, and the median \textbf{4p}$_{PQP2}$ CARMENES data RV zero-point value was been applied in order to more easily compare both datasets with the model expectations. Top: mean model prediction (gray solid line), together with contemporaneous HARPS-N (black), CARMENES (red), and HIRES (purple) RVs overplotted. Bottom: model residuals, together with 1- and 2-$\sigma$ GP uncertainty bands (shaded dark and light grey regions, respectively). Takeaway: The preferred SM21 model is overfitting to the HARPS-N data, which can be seen in the increased spread about the residual=0 line for both HIRES and CARMENES data during epochs with contemporaneous HARPS-N data.}
    \label{fig:prediction}
\end{figure*}

\begin{figure*}
    \centering
    \hspace*{-0.5in}
    \includegraphics{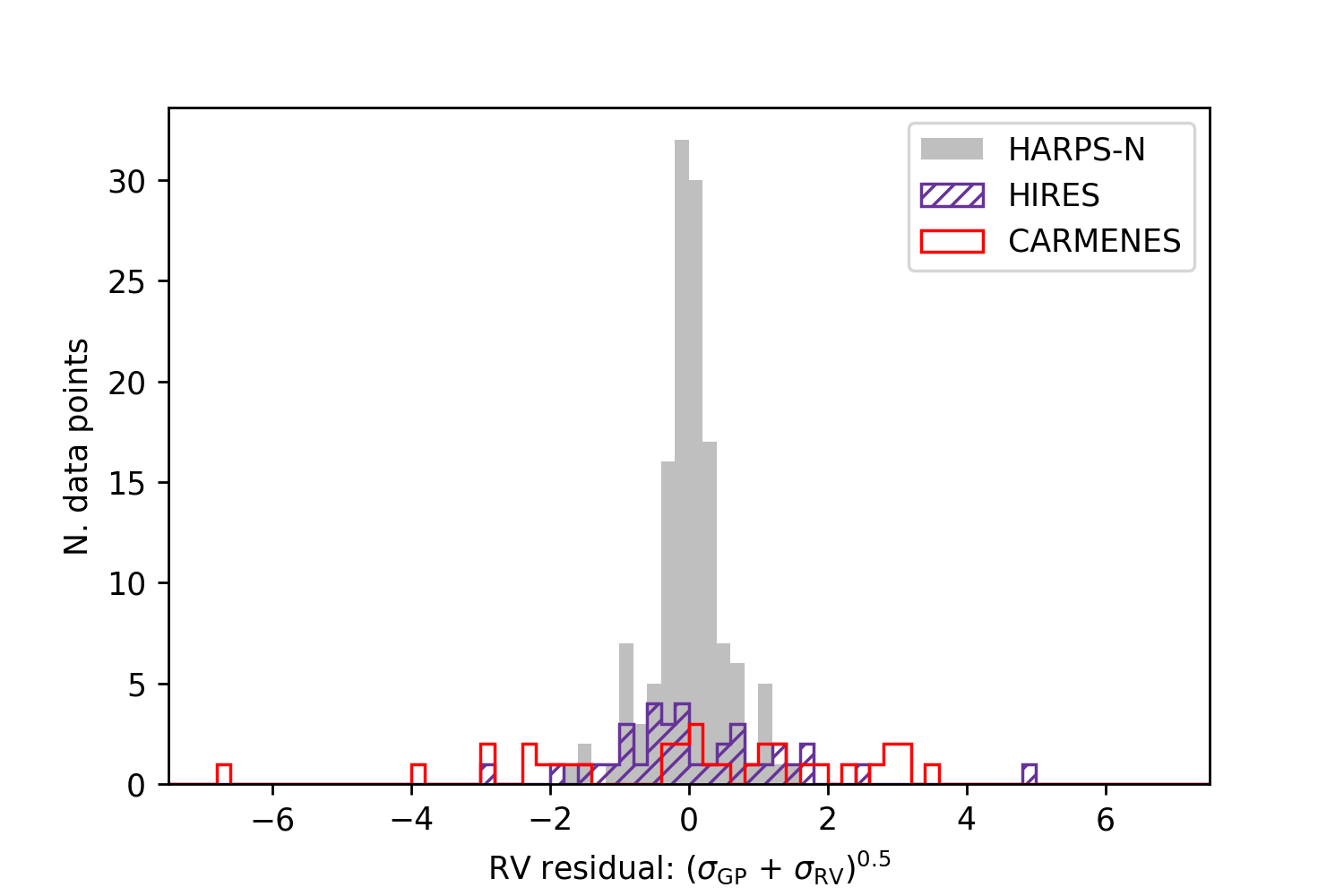}
    \caption{Another visualization of Figure \ref{fig:prediction}. Histograms of the RV residuals, given in units of standard deviations from the mean prediction. Takeaway: The broader and more uniform distribution of HIRES and CARMENES residuals relative to the HARPS residuals is another hallmark of overfitting.}
    \label{fig:residuals}
\end{figure*}

Our goal was to test the predictiveness of the preferred SM21 model\footnote{conditioned on the HARPS-N data of SM21; see Section \ref{sec:corr_data}.} using CV. In an ideal situation, we would do this by evaluating the model's performance on new HARPS-N data, unseen by the trained model. Lacking this, we constructed two ad hoc ``validation sets:'' a timeseries of Keck/HIRES data contemporaneous with the SM21 HARPS-N data, and the CARMENES data presented in SM21 (that the model was also trained on, but which were treated as independent from the HARPS-N data; see section \ref{sec:why}). By chance, this results in a nearly perfect 80\%/20\% split for both validation sets (80.3\%/19.7\% for HARPS-N/CARMENES, and 78.9\%/21.1\% for HARPS-N/HIRES). In Figures \ref{fig:prediction} and \ref{fig:residuals}, we show two visualizations of the results of performing CV on these two validation sets. Figure \ref{fig:prediction} shows the GP prediction of the SM21 preferred model, together with the HARPS-N data on which it was trained and conditioned. The contemporaneous HIRES data and CARMENES data and their residuals are overplotted. Figure \ref{fig:residuals} shows the residuals of this fit, given in terms of standard deviations from the mean GP prediction. In both figures, the residuals of the HIRES and CARMENES data have a much wider spread about 0 than the HARPS-N points. Because our intention was to evaluate the existing model, we did not re-train the GP hyperparameters in order to compute the prediction shown in Figure \ref{fig:prediction}. Rather, we used the median parameters of the SM21 \textbf{4p}$_{PQP2}$ model, conditioned on the HARPS-N data published in that study, to predict RV values at each of the CARMENES and HIRES epochs. 

Our interpretation of the difference in residual distributions shown in these two figures is that the preferred SM21 model fits data included in its training set (i.e., the HARPS-N data) significantly better than \textit{contemporaneous} data not included. In other words, the model is not predictive. This is a hallmark of overfitting, and indicates that the preferred SM21 model is not representative of the process generating the data.

An important counter-interpretation is that the V1298~Tau RVs measured by HARPS-N, HIRES, and CARMENES show different activity signals, and not that the preferred SM21 model is overfitting. In particular, starspots cooler than the stellar photosphere cause RVs collected in redder bands, where the contrast between spot and photosphere is lower, to show lower variability amplitudes \citep[e.g.,][]{Carpenter_NIRPhotometricVar_2001AJ, Prato_VisIR_2008ApJ, mahmud_starspot-induced_2011}. In addition, we expect different instruments to have different RV zero-point offsets. Importantly, these two effects cannot explain the increased out-of-sample residual spread observed in Figures \ref{fig:prediction} and \ref{fig:residuals}\footnote{assuming that stellar activity signals observed by different instruments can be described as linear combinations; see Section \ref{sec:corr_data}.}; the preferred SM21 model fitted the CARMENES zero-point offset, white noise jitter value, and activity amplitude, and those values have been applied to the CARMENES data. To account for the potential differences between the HIRES and HARPS-N RVs, we applied an RV zero-point offset and scale factor (0.76) that minimizes the residual spread (i.e., we applied a best-fit linear model to the HIRES data in order to minimize $\chi^2$ with respect to the GP model prediction). See section \ref{sec:corr_data} for further discussion of this point.

\begin{figure*}
    \centering
    \hspace*{-0.5in}
    \includegraphics{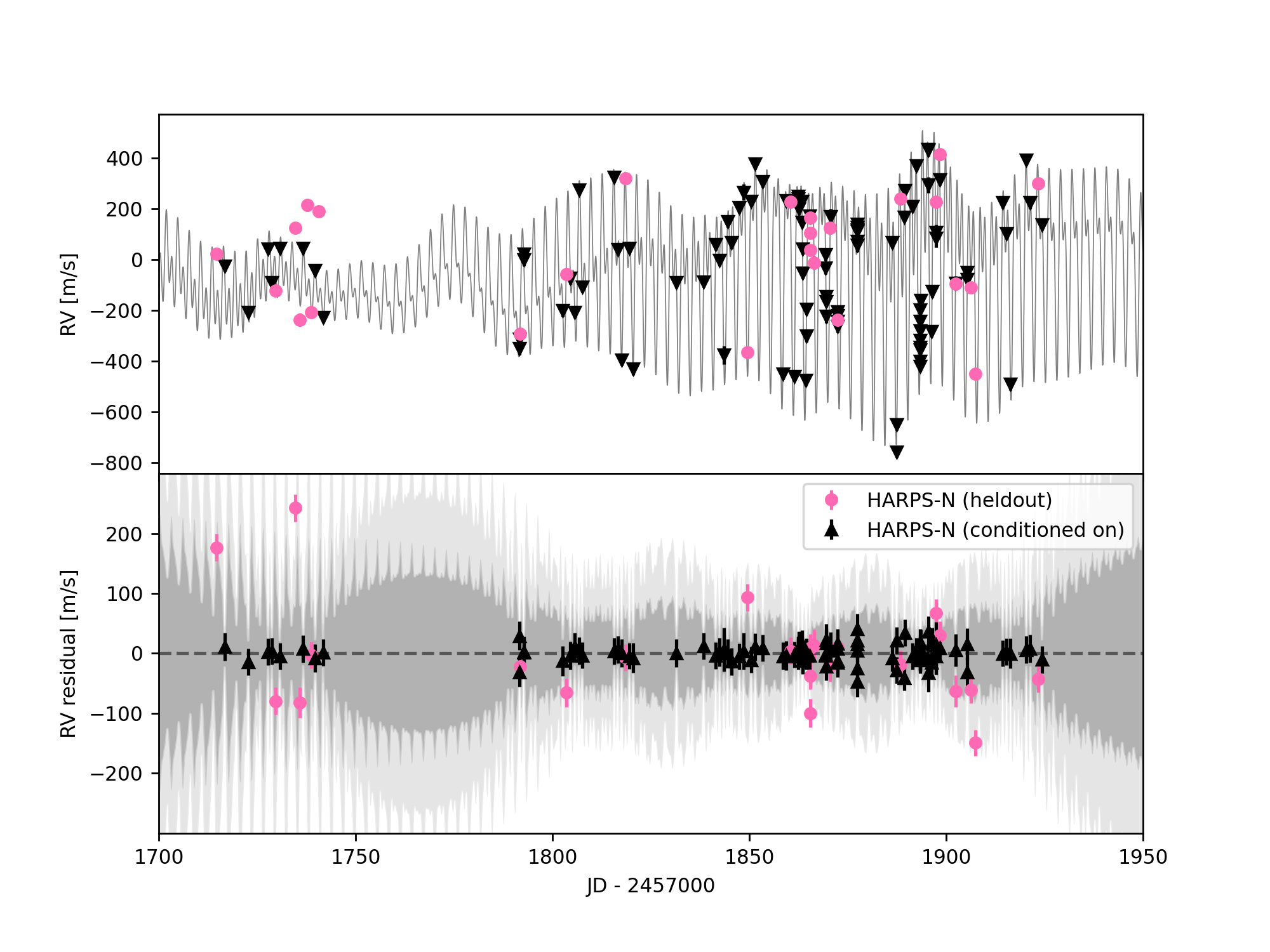}
    \caption{Same as Figure \ref{fig:prediction}, except that the model prediction is computed by conditioning on a randomly-selected 80\% subset of the HARPS-N data, as described in the text, as the residuals are computed for the 20\% subset that was held-out. Takeaway: The effect seen in Figure \ref{fig:prediction} cannot be explained by instrument- or wavelength-dependent systematics, because the same larger residuals are seen within the data taken by only HARPS-N.}
    \label{fig:harps_prediction}
\end{figure*}

\begin{figure*}
    \centering
    \hspace*{-0.5in}
    \includegraphics{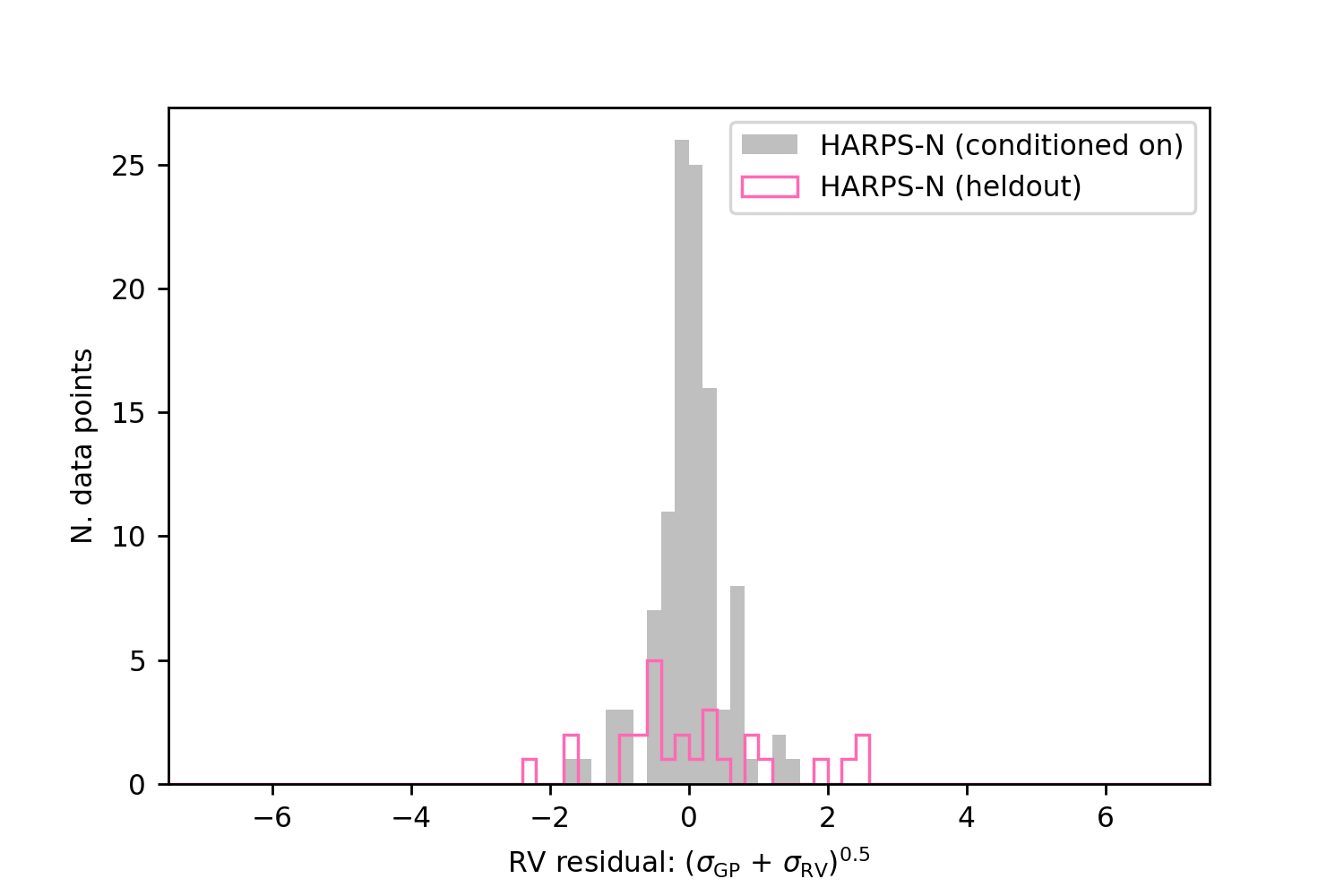}
    \caption{Another visualization of Figure \ref{fig:harps_prediction}. Same as Figure \ref{fig:residuals}, except computed using the same method as for \ref{fig:harps_prediction}. Takeaway: the larger and more uniform spread of residuals for HARPS-N data on which the model was conditioned provides more evidence that the preferred SM21 model is overfitting.}
    \label{fig:harps_residuals}
\end{figure*}

Another potential explanation for the phenomenon observed in Figures \ref{fig:prediction} and \ref{fig:residuals} is that the activity signals observed by HARPS-N, CARMENES, and HIRES are fundamentally different; i.e., the signal observed by one instrument is not a linear combination of the signal observed by another. This might occur because, for example, all three instruments have $\sim$km/s instrumental systematics relative to one another, or because the shape of the activity signal changes significantly with wavelength. To rule out this explanation and provide more evidence that the effect we're seeing is actually overfitting, and not instrument-specific differences, we repeated the experiment above using only HARPS-N data. We randomly selected 80\% of the HARPS-N data published in SM21, conditioned the preferred SM21 model on that subset, and computed the residuals for the random ``held-out'' 20\%. The results are shown in Figures \ref{fig:harps_prediction} and \ref{fig:harps_residuals}. Even though these held-out 20\% were included in the training process (i.e., they informed the values of the hyperparameters), we observed substantially larger residuals than for the conditioned-on subset. This experiment provides additional evidence for overfitting, and not instrumental- or wavelength-dependent systematics. 

It is worth noting that we distinguish between residual distributions (Figures \ref{fig:harps_prediction} and \ref{fig:residuals}) ``by-eye'' in this paper, but this technique will not generalize for more similar residual distributions. Residual diagnostic tests (see \citealt{Caceres:2019a} for an example) will be helpful in generalizing this methodology.

\section{Potential Causes of Overfitting}
\label{sec:why}

This section points out several potential causes of the overfitting described in the previous section, and advises on how to detect and/or ameliorate these effects. We do not attempt to quantify the effect of each of these on the overfitting discussed in the previous section, but intend this as a qualitative discussion. Many of these effects are potentially relevant for stars other than V1298~Tau. 

Importantly, this is not a list of ``mistakes,'' but a list of assumptions we questioned throughout the process of trying to explain why the preferred SM21 fit was overfitting. We encourage future close investigation of each of these phenomena, both for V1298~Tau and other objects. This list is not exhaustive.

\subsection{Correlated Datasets vs Datasets that Share Hyperparameters}
\label{sec:corr_data}

The mathematical formalism in this section is essentially identical to that of \citet[see their section 3.2]{Cale:2021aa}, but was developed independently. We encourage readers to compare our explanations, and we ask readers to also cite \citet{Cale:2021aa} whenever referencing Section \ref{sec:corr_data} of this paper.

There is a difference between correlated measurements that are allowed to have different GP amplitudes and datasets that share GP hyperparameters but are themselves uncorrelated. We are motivated to stress this distinction by the need in RV timeseries fitting to write down the joint likelihood of a model applied to datasets taken from several different instruments. As a concrete example, let's consider three fictional RV data points, the first two from HIRES and the next one from CARMENES, to which we would like to fit a GP model. Because of the different bandpasses of HIRES and CARMENES, we might expect the same stellar activity signal to have a different amplitude when observed by these two instruments. However, we might expect the time-characteristics of the signals to be identical. In other words, we expect the CARMENES activity signal to be a scalar multiple of the HIRES activity signal.\footnote{With different RV offsets as well, so technically a linear combination, not just a scalar multiple.} As discussed in Section \ref{sec:overfitting}, this assumption is borne out, at least to first order, in observations of other active stars at different wavelengths (see, e.g., \citealt{mahmud_starspot-induced_2011}, who investigated the RV activity of the T-Tauri object Hubble I 4 with contemporaneous infrared and optical spectra taken with different instruments), but this point warrants further scrutiny. Comparing the variability of active stars with different instruments, as well as the variability of the sun with different solar instruments, is an important endeavor.

Another important caveat is the use of different techniques for computing RVs from stellar spectra (e.g., the iodine/forward-modeling technique of HIRES vs simultaneous reference/CCF technique of CARMENES and HARPS-N). Switching from one of these techniques to another is not expected to affect an astronomer's ability to recover common Keplerian signals, but spot activity is not a simple Doppler shift. More work is needed to understand and model spot activity at the spectral level. We proceed by assuming that modeling the same spectrum using an iodine/forward-model and with a simultaneous ThAr lamp reference (as an example) will only change the effective wavelength range of the spectrum that is used to compute RV, and therefore affect only the amplitude of spot-induced variations.

Assuming linearly-related GPs for different instruments, we can write down the joint covariance matrix for our three fictional data points, allowing unique amplitude terms $a_{\rm C}$ and $a_{\rm H}$ for each dataset, and assuming an arbitrary kernel function k$_{i,j}$ describing the covariance between RVs at times t$_i$ and t$_j$:

\begin{equation}
    C_{\rm joint} = \begin{pmatrix}
    a_{\rm H}^2 k_{0,0} & a_{\rm H}^2 k_{0,1}& a_{\rm H} a_{\rm C} k_{0,2} \\
    a_{\rm H}^2 k_{1,0} & a_{\rm H}^2 k_{1,1}& a_{\rm H} a_{\rm C} k_{1,2} \\
    a_{\rm C} a_{\rm H} k_{2,0} & a_{\rm C} a_{\rm H} k_{2,1}& a_{\rm C}^2 k_{2,2} \\
    \end{pmatrix}.
\end{equation} Optimizing the hyperparameters of a fit that uses this covariance matrix to define the GP likelihood will give the desired result. 

SM21, following many other fits in the literature, constructed an independent covariance matrix for each RV instrument in their dataset and summed the log(likelihoods) given by these together. This allows each RV dataset to be independent; i.e., a datapoint taken by HIRES is not correlated with a datapoint taken at exactly the same time by CARMENES. Figures \ref{fig:radvel_indep} and \ref{fig:radvel_mult} illustrate the difference between these two likelihood definitions for data for a different object (chosen because it is easier to see the effect using this dataset).

This assumption of independent data for each instrument effectively adds additional free parameters to a model, and makes it more susceptible to overfitting. This is also why, in Figures \ref{fig:prediction} and \ref{fig:residuals}, we could demonstrate that the preferred SM21 model was overfitting by comparing the model prediction conditioned on HARPS-N data to the CARMENES data; the CARMENES data influenced the final values of the hyperparameters, since they were shared between the two Gaussian processes, but otherwise the datasets were treated as independent.

To illustrate the effects discussed in this paper, we used a modified version of \texttt{radvel} \citep{Fulton:2016aa}, built on \texttt{tinygp} \citep{dan_foreman_mackey_2022_7269074}, that treats the models for different instruments as correlated, but allows each instrument its own GP amplitude, white noise jitter term, and RV zero-point offset term.\footnote{This is slightly different from the GP prescription in \texttt{juliet} \citep{juliet}, which does not allow different amplitudes for individual RV instruments.} The difference between the previous version of \texttt{radvel} and this modified version is also illustrated in Figures \ref{fig:radvel_indep} and \ref{fig:radvel_mult} in the Appendix. This modified version of the code is available at \href{https://github.com/California-Planet-Search/radvel/tree/tinygp}{https://github.com/California-Planet-Search/radvel/tree/tinygp}.

Future work should continue to test this assumption by obtaining simultaneous (or near simultaneous) RVs for a variety of stellar types with different instruments, across a wide range of bandpasses. 

\subsection{P$_{\rm rot}$ and P$_{\rm rot}$/2}
\label{sec:p2}

Another practice that may have made the SM21 preferred fit susceptible to overfitting involves constructing a GP kernel with one term at the rotation period and another term at its first harmonic. In other words, the SM21 preferred model kernel has the following form:

\begin{equation}
\label{eq:perover2}
    C_{ij} = f_1(P_{\rm rot}) +  f_2(P_{\rm rot} / 2),
\end{equation}

\begin{figure*}
    \hspace*{-0.5in}
    \vspace*{-0.9in}
    \centering
    \includegraphics[trim={0 0 0 1in},clip]{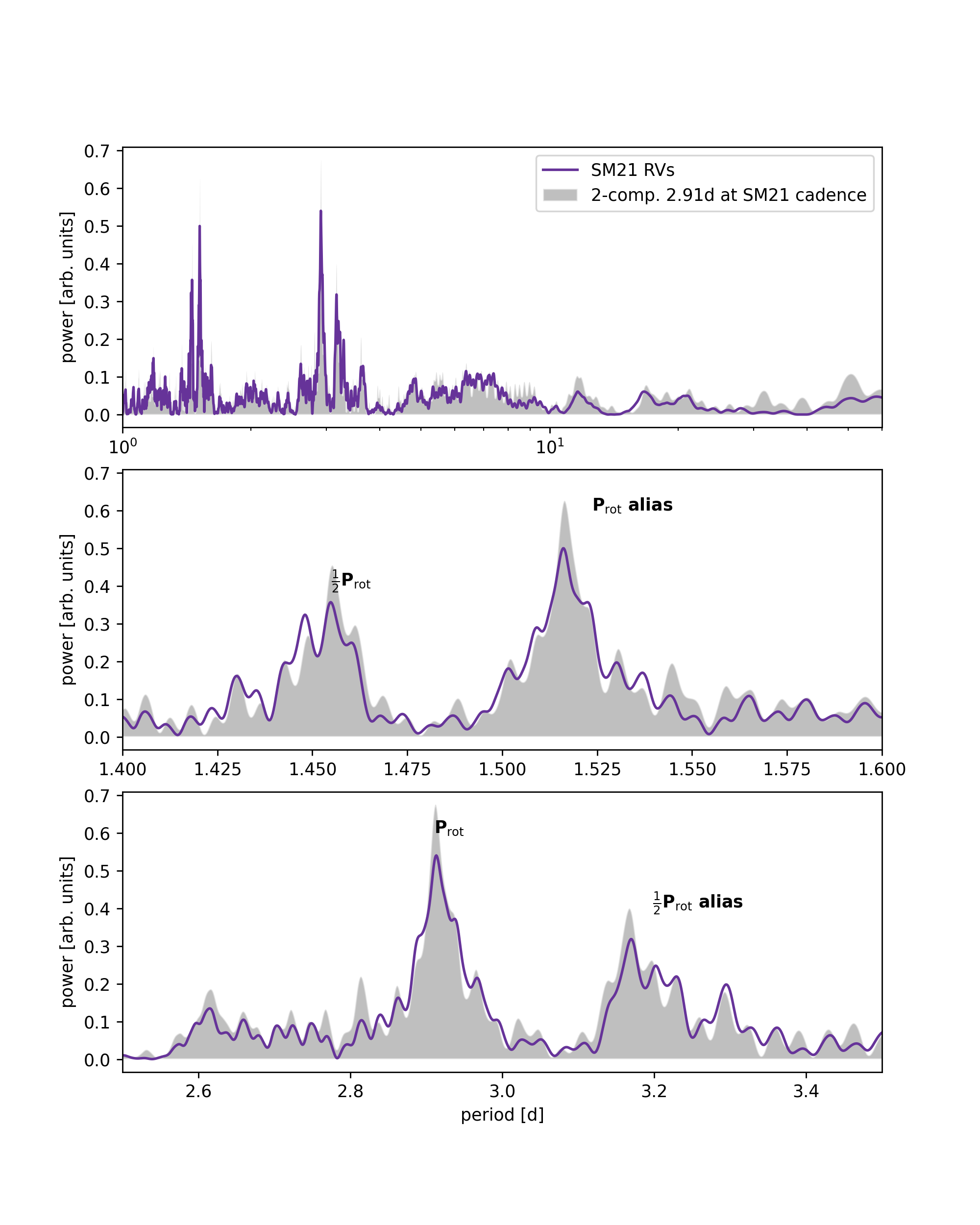}
    \caption{Lomb-scargle periodogram of all RV data presented in SM21, and 2-component sinusoidal fit passed through the same window function. Top: Periodogram of all RVs (solid purple line) and a 2-component sinusoidal fit to the data (filled grey). Middle/bottom: same, but zoomed in. The rotation period, its harmonic, and its 1d aliases are labeled. Takeaway: the dominant Lomb-Scargle periodogram structure can be explained as harmonics and aliases of a single period at 2.91d.}
    \label{fig:rvs_ls}
\end{figure*}

\begin{figure*}
    \centering
    \hspace*{-0.5in}
    \includegraphics{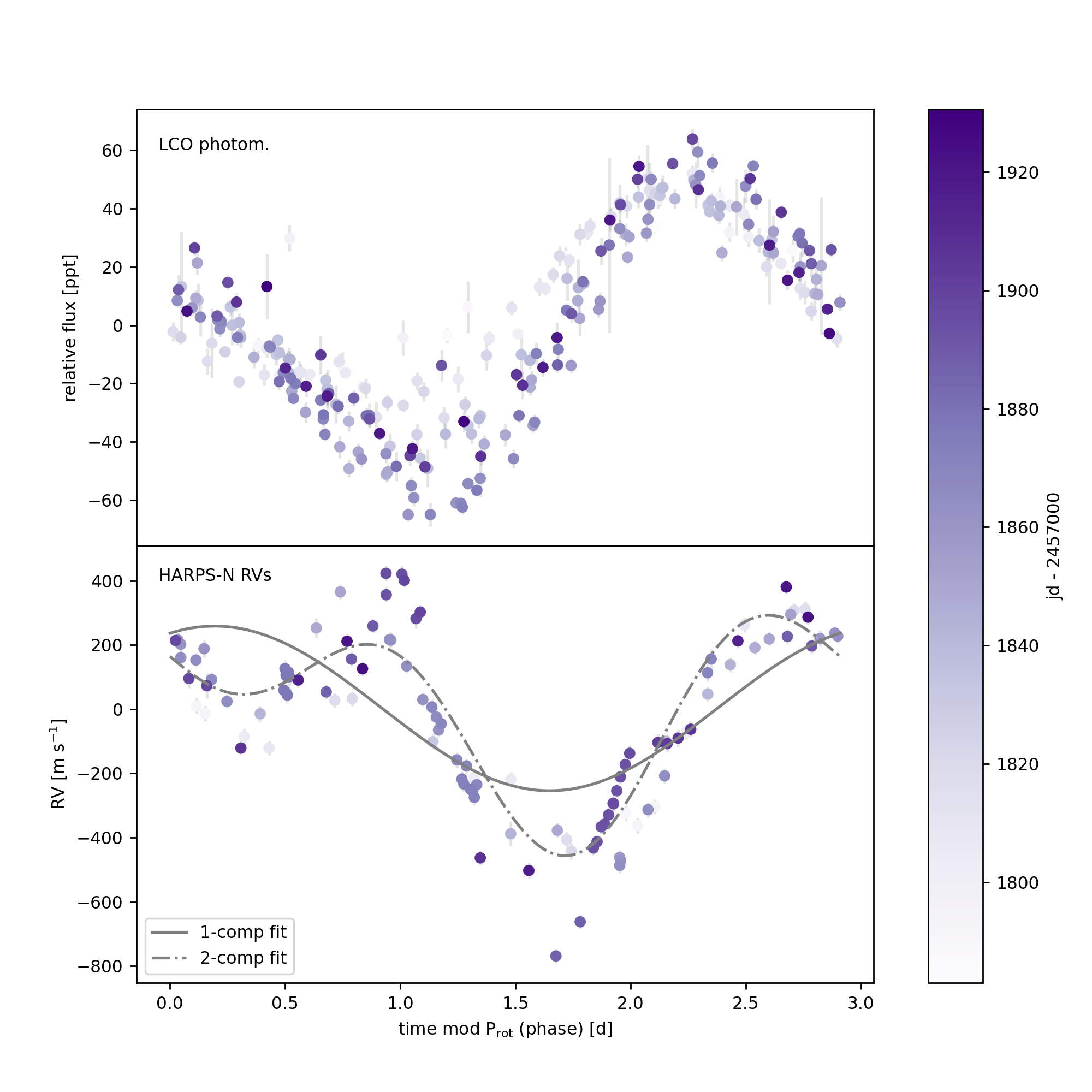}
    \caption{HARPS-N RVs and contemporaneous LCO photometry from SM21, phase-folded to the rotation period and colored by observation time. Top: LCO photometry. Bottom: HARPS-N RVs, with fitted jitter values from the preferred SM21 fit added to the error bars. 1- and 2-component sinusoidal fits are also shown. Takeaway: the presence of a strong periodogram peak at P$_{\rm rot}$/2 results from the higher-order shape of the RV rotation pattern. This pattern is not present in the LCO photometry, which is approximately sinusoidal over the rotation period.}
    \label{fig:rv_phasefold}
\end{figure*}

To understand the motivation for this, we first need to scrutinize the RV signal in Fourier space. Figure \ref{fig:rvs_ls} shows the Lomb-Scargle periodogram of all RV data presented in SM21, zooming in on two important parts of period space. There are four extremely significant peaks in the RVs, which can all be explained with a single periodic signal at 2.91d, the rotation period identified by SM21. Along with a strong peak at 2.91d (hereafter P$_{\rm rot}$), there is a signal at P$_{\rm rot}$/2, which is often observed in RVs of stars showing starspot-induced variability \citep{Nava:2020aa}. The other two strongly significant peaks can be explained as 1-day aliases of P$_{\rm rot}$ and P$_{\rm rot}$/2. In other words, the dominant RV signal is periodic, but requires a two-component sinusoidal fit (i.e., it needs more terms in its Fourier expansion) in order for the fit to reproduce the shape of the curve. This is visualized in Figure \ref{fig:rv_phasefold}, which shows the RVs phase-folded to P$_{\rm rot}$. In summary, the RV curve comprises a single periodic pattern, but that pattern is not a simple sinusoid.


The preferred SM21 model kernel sums two approximately quasi-periodic terms, one at P$_{\rm rot}$ and one at P$_{\rm rot}$/2, because the approximate quasi-periodic kernel used in SM21 (SM21 equation 1; derived in \citealt{celerite}) is less flexible than the standard quasi-periodic kernel (SM21 equation 3). In other words, the approximate kernel is less capable of fitting non-sinusoidal shapes. However, each term was modeled with its own independent exponential decay timescale. This adds an additional free parameter to the fit, which exacerbates the potential for overfitting.

The most straightforward way to address this is to construct a model with fewer unnecessary free parameters, for example by equating the parameters L$_1$ and L$_2$ in SM21 equation 1. A more complicated suggestion, which would be an excellent avenue for further study, is to leverage the correlation between the photometry and RVs, following, for example, \citet{Rajpaul:2015aa}. This requires assuming (or fitting for) a relationship between a photometric datapoint and an RV datapoint at the same time. Our preliminary investigations along these lines indicate that the FF' formalism, which models an RV signal as a function of a simultaneous photometric (F) dataset and the time derivative of the photometric dataset (F'; \citealt{Aigrain:2012aa}), does not allow for a good phenomenological match between the LCO photometry and the contemporaneous RVs, but the derivative of the LCO photometry appears to fit better (i.e., the RV curve appears to be possible to model as a linear combination of the F' component only)\footnote{This was also noted in SM21.}. Future work could write down a joint GP formalism that models RVs as the time derivative of the photometry (such a formalism would be very similar to that of \citealt{Rajpaul:2015aa}). 

Regardless, in order to be confident in the relationship between the photometry and the RVs, as well as to pick out the components of the RV that do not occur at P$_{\rm rot}$, we suggest a very high-cadence (several observations per night) RV follow-up campaign with contemporaneous photometry\footnote{As of 1-30-23, V1298 Tau will unfortunately not be reobserved with TESS through year 6. We used \texttt{tess-point} \citep{Burke:2020aa} to make this determination.} in order to develop a high-fidelity model of the stellar variability\footnote{It is worth pointing out that similar strategies have been successful before, e.g., to measure the mass of Kepler-78 b (\citealt{Pepe:2013aa}, \citealt{Howard:2010aa})}. It is important to note that this campaign need not be performed by an RV instrument with 30 cm s$^{-1}$ precision; \citet{Johnson:2022aa} demonstrated 6-7 \ms{} RMS precision with HIRES over several hours, even though the stars moves by hundreds of \ms{} over even a single night. This level of instrumental RV error should be sufficient to understand the stellar activity, so long as the cadence is as high as possible.

\subsection{Keplerian Parameters Enable Overfitting in the Presence of Un-modeled Noise}

A Keplerian signal has five free parameters (semi-amplitude, eccentricity, argument of periastron, time of periastron, and period). A model with two Keplerian signals therefore has 10 additional free parameters than a model without. To first order, more free parameters means more model flexibility. This problem can be addressed using model comparison, which penalizes complexity. However, if there is un-modeled noise in the data, including additional Keplerian signals in the model can lead to overfitting; for example, high eccentricity Keplerian models have similar properties to delta functions, which have relatively ``flat'' RV curves, except for a spike in RV near periastron. With insufficient sampling, outlier data points can be overfit with eccentric Keplerian signals.

\begin{figure*}
    \hspace*{-0.5in}
    \centering
    \includegraphics{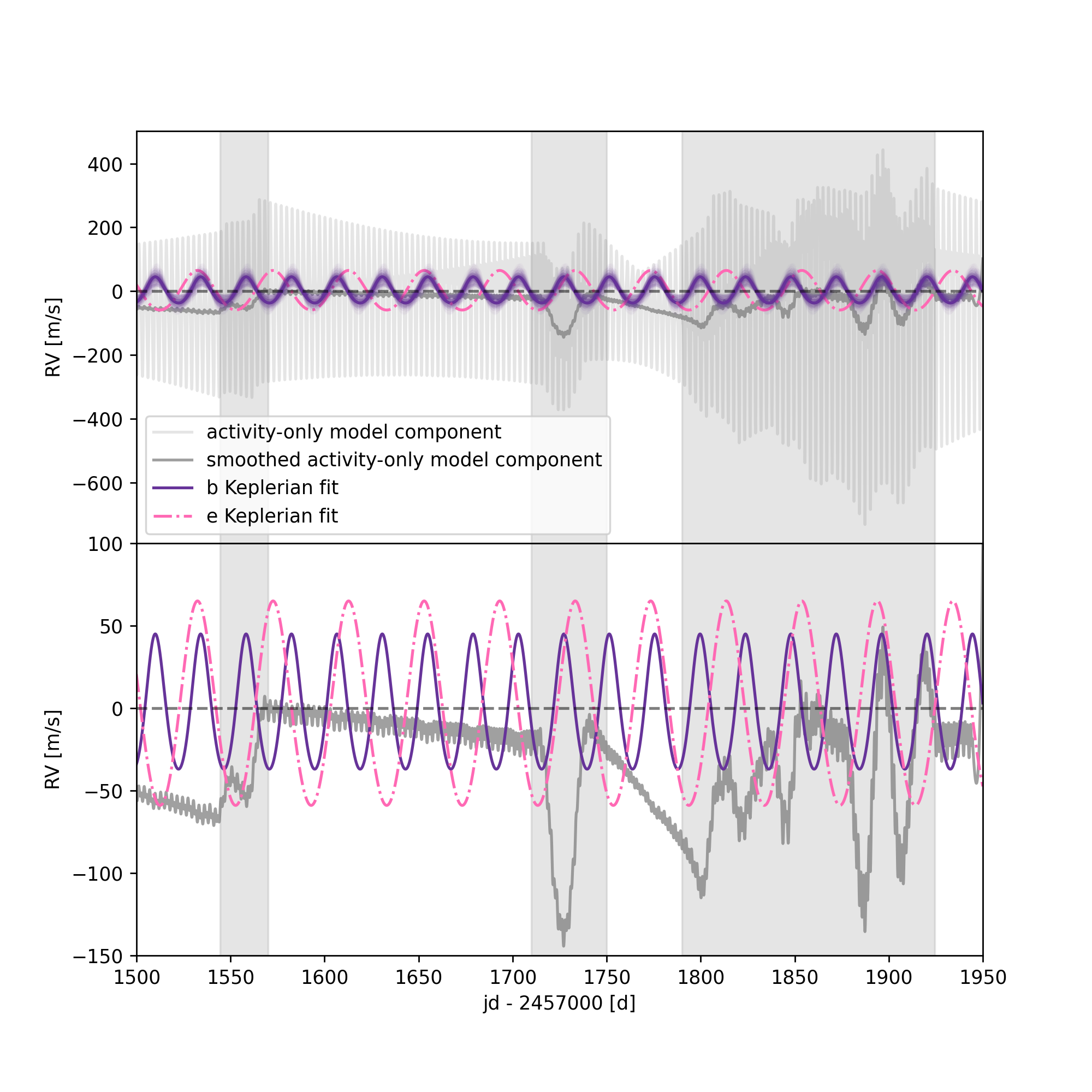}
    \caption{Smoothed activity-only component of the preferred model of SM21, together with the Keplerian model components. Top: 100 random draws from the posterior describing the planet b Keplerian are also shown, to illustrate that this effect holds true across the posterior, and not simply for one point estimate. The light gray solid line shows the full activity-only model component, and the darker grey shows this model averaged over a (randomly chosen) 11.2 d timescale. (Note that the same pattern holds when choosing a slightly different smoothing timescale; i.e., this is not a result of aliasing.) Shaded grey regions indicate where there are observations. Bottom: same as top, but with a zoomed-in y axis. Takeaways: the activity-only component changes suddenly in windows of time where there are observations. When the activity-only component is averaged over shorter-timescale variations, the GP contributes to the fit on timescales similar to the Keplerians, even interfering destructively at some times. This casts doubt on the reality of the Keplerian signals reported in SM21, indicating that they may be favored because of overfitting.}
    \label{fig:keplerian_destruction}
\end{figure*}

A common worry in the RV modeling community is that using GPR to model stellar activity will ``soak up'' Keplerian signals, leading to underestimates of Keplerian RV semi-amplitudes (discussed in \citealt{Aigrain:2022aa}), even when modeled jointly. However, we find evidence for the opposite effect in the SM21 preferred fit: that the Keplerian signals function as extra parameters that make the model susceptible to overfitting, and the GP is forced to compensate. Examining Figure \ref{fig:keplerian_destruction}, which shows the contributions to the mean model prediction from the Keplerians and the activity-only portion of the mean GP model\footnote{The activity-only portion is isolated following SM21, subtracting the Keplerian mean model from the total mean GP prediction.}, we find that the activity model interferes with the Keplerian model where RV data exists. This is seen most readily when smoothing the activity model over several rotation periods (effectively low-pass filtering the activity model). 

We can explain this behavior by imagining that there is some un-modeled noise source in the data that is inconsistent with Keplerian motion or quasi-periodic variability (see next section). If some non-physical combination of parameters fits the data better at an epoch with many data points that is affected by this noise source, this may outweigh the negative Bayesian evidence contributions from 1) the added complexity and 2) the worse fit at epochs with fewer data points. We would then expect the Keplerian model to \textit{oversubtract} at epochs with fewer data points (e.g., around jd = 1725 in Figure \ref{fig:keplerian_destruction}).

This effect suggests that the Keplerian signals in the SM21 preferred fit are not a viable description of the RV variability at timescales greater than the rotation period. More effort certainly needs to be spent understanding this phenomenon, but in the meantime we suggest performing CV tests in order to detect overfitting of this nature.

\subsection{Differential Rotation}
\label{sec:diffrot}

The previous subsections all argue that the preferred SM21 fit had too many free parameters (or effective free parameters) that allowed the model to overfit. In other words, we have argued that a simpler model (one for which the GP predictions for each instrument are scalar multiples of each other, a single period is present in the kernel, and no Keplerian signals are present in the model) would be more predicitive, albiet perhaps with larger uncertainties. In this section, we suggest that this much simpler proposed model is still insufficient, because the host star has multiple, differentially rotating, active regions. 

Differential rotation may not be the un-modeled noise source that we propose is affecting the SM21 preferred fit. The conclusions of this paper do not change if this is true. We discuss it here because it is potentially widely relevant, especially for young stars. We call for more work on modeling and understanding differential rotation in RVs. 

\subsubsection{Evidence for a Strong Differential Rotation Signal from Photometry}

\begin{figure*}
    \hspace*{-0.5in}
    \vspace*{-0.5in}
    \centering
    \includegraphics{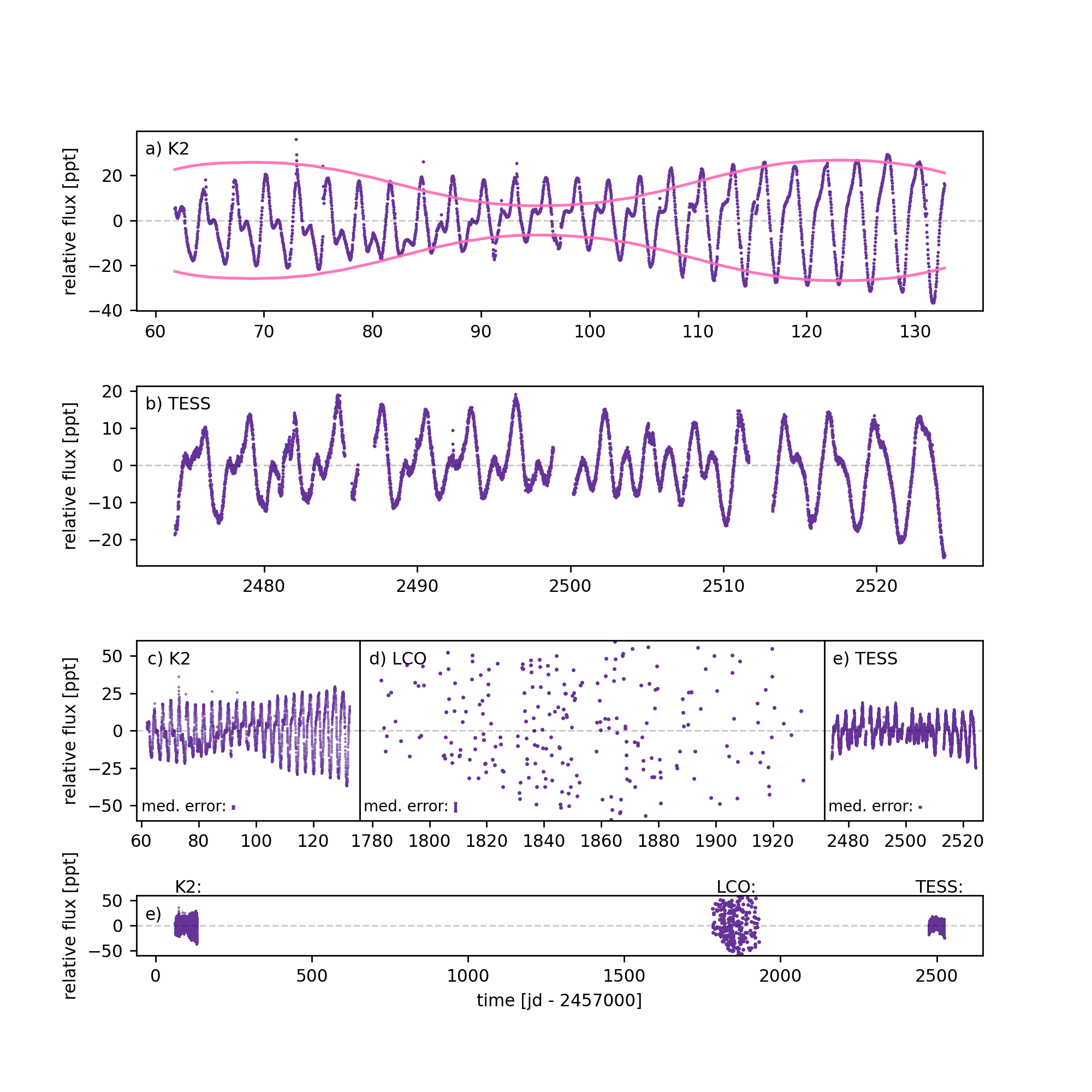}
    \caption{A tour of the relevant photometry of the star V1298 Tau. Panel a: detailed view of the K2 photometry (purple points), with a beating envelope over-plotted in solid pink. The beating envelope is drawn to illustrate the effect of spot beating on overall variability amplitude, not to precisely fit the data. The envelope drawn is constructed from the beating of three sinusoids at 2.70, 2.85, and 3.00 d. Signatures of beating can be seen by eye: two peaks of different amplitudes phase up toward the end of the K2 baseline, producing a single-peaked variability pattern and a larger overall variability amplitude. Panel b: detailed view of the TESS photometry (purple points). Beating characteristics are also visible, although the baseline is shorter than that of K2. Panels c, d, and e: relative views of K2, LCO, and TESS photometry, emphasizing relative time baseline and variability amplitude. A typical error bar for each dataset is also shown in the bottom left corner of each panel. The differences in wavelength coverage and flux dilution between the K2, LCO, and TESS photometry largely account for the overall differences in amplitude of the signals. Both the K2 and TESS data cover less than one complete beat period of the two largest-amplitude periodic signals, but the LCO photometry (which is contemporaneous with the RVs of SM21) covers a longer time baseline. Panel e: All photometry, plotted on the same panel to emphasize relative time elapsed between each dataset. Takeaway: differential rotation effects are visible by eye in both the K2 and TESS datasets.}
    \label{fig:phot}
\end{figure*}

In the K2 and TESS photometry of V1298 Tau (Figure \ref{fig:phot}), two periodic signals of different amplitudes are visible by eye. These peaks are coherent in phase towards the end of both baselines, producing a larger overall photometric variability amplitude. Although each baseline covers only a portion of the beat periods implied by these different periods coming into and out of phase, the beating ``envelope'' is still easily distinguished. To guide the eye, we over-plotted the shape of the beating envelope formed by the three dominant periods in the Lomb-Scargle periodogram of the K2 data. 

Multiple closely-related periodicities are also apparent in the periodograms of the K2 and TESS data (and the LCO data, albeit at lower significance, potentially due to the lower cadence of that dataset; Figure \ref{fig:photom_periodogram}). In particular, over both the K2 and TESS baselines, a dominant periodicity at 2.85 and 2.92 d, respectively, and two less prominent periodicities (one at a larger period, and one at a smaller period) are present. The multiple periodicities in the light curve, visible both in the shape of the beating envelope and in Fourier space, have often been interpreted as a smoking gun of differential rotation (see, e.g., \citealt{Lanza1994aa}, \citealt{Frasca2011aa}). It is important to note, however, that short spot lifetimes may also produce the observed photometric pattern, and have been shown in simulations to be easily confused with differential rotation (see, e.g., \citealt{Basri:2020aa}). Longer photometric time baselines than are available in the photometric data presented in this paper are needed to distinguish between the two. The conclusion of this section (that there is a noise source visible in photometry that is un-modeled in the SM21 preferred model) would remain unchanged in this case, but this interpretation has important implications for future modeling efforts. That the signals arise from a close binary is ruled out by the multiple nearby periods in the light curve (rotation of two tidally extended binary stars can produce a similar pattern, but with a single period), while astroseismic pulsations are ruled out by the amplitude and period of the variability; V1298 Tau is a PMS 1.2M$_{\odot}$ star with $\log{g}$=4.48 (SM21), which we would expect to be oscillating on the scales of minutes and $\lesssim$1 ppt, not days and 20 ppt (\citealt{Chaplin2013aa}; see their Figure 3). 

\subsubsection{Effect on RVs}

Assuming that V1298 Tau is differentially rotating, it is possible that the combination of a multiply periodic structure with insufficient cadence is leading the GP to prefer a more complex model. In other words, the data is not consistent with a quasi-periodic structure, so a simple quasi-periodic model will not be preferred over a more complex model (e.g., one with Keplerians at longer periods), even if neither is predictive. Even a secondary active region with 5\% the RV amplitude of the primary structure (reasonable given the photometric amplitude ratios) would incur an RV variability of 20 m/s, significantly greater than the instrumental floor of HARPS-N, CARMENES, and HIRES. 

An important clarification is that this conclusion is consistent with the discussion in Section \ref{sec:p2}. Although there is a clear periodic 2.91 d signal visible in Figure \ref{fig:rv_phasefold}, there is also $\sim$200 \ms{} of scatter around this signal.  It is possible that this scatter may contain coherent signals at other periods that are unresolvable with the current RV cadence.

\begin{figure*}
    \hspace*{-0.5in}
    \centering
    \includegraphics{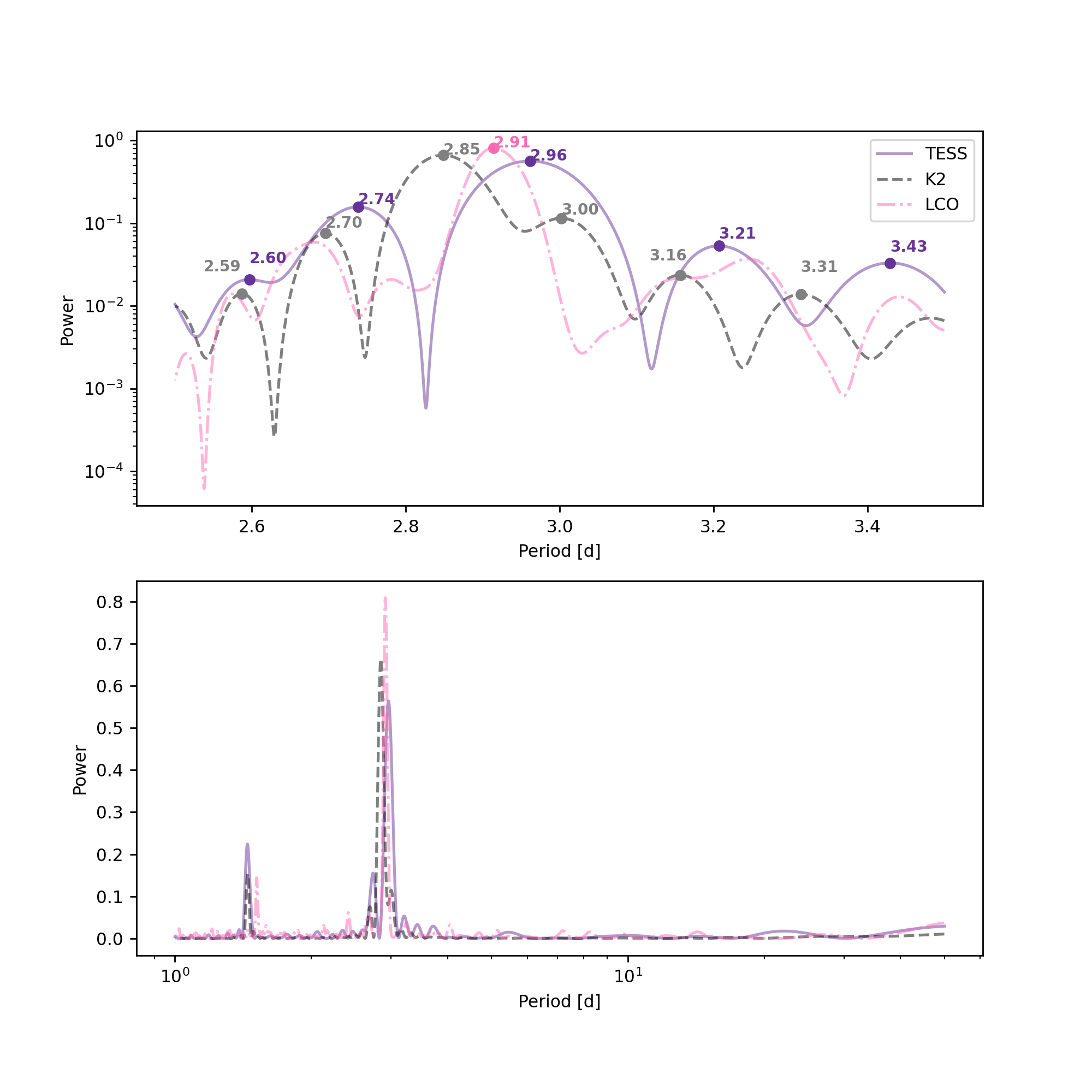}
    \caption{Lomb-Scargle periodograms of the photometric data shown in Figure \ref{fig:phot}. Top: Zoom-in on the presumed rotation period, showing several nearby peaks in all three datasets. Bottom: Same as top over a wider period range. Takeaway: multiple closely-related periodicities are visible in Fourier space for all three photometric datasets, more evidence for differentially rotating active regions.}
    \label{fig:photom_periodogram}
\end{figure*}

Complicating this already complicated story is the fact that the dominant periodicity appears to change over time (Figure \ref{fig:photom_periodogram}). This provides further motivation for our major recommendation, first given in Section \ref{sec:p2}: V1298 Tau appears to be a multiply-periodic star with evolving periodicity. A high-cadence (several data points per night) RV campaign is necessary to construct a high-fidelity activity model. The high cadence is necessary to resolve the close periodicities due to apparent differential rotation. Care should be taken to ensure that the periods do not evolve significantly over the observing baseline, or that this effect is sufficiently modeled.

\section{Summary \& Discussion}
\label{sec:discuss}

In this study, we have presented evidence that the preferred model of SM21 is overfitting using two ad hoc ``validation'' data sets: one set of contemporaneous HIRES and CARMENES data, and one set of artificially held-out HARPS-N data. The effects that we have proposed may be responsible for the non-predictiveness of the preferred SM21 model are:

\begin{itemize}
    \item The RV datasets from different instruments are treated as uncorrelated, allowing the model more freedom.
    \item The SM21 preferred model includes two summed quasi-periodic terms at P$_{\rm rot}$ and P$_{\rm rot}$/2 in their kernel, each with its own free exponential decay parameter. This additional free parameter grants the model unnecessary flexibility.
    \item The SM21 model also includes parameters describing eccentric Keplerian signals, which grant even more degrees of freedom. 
    \item We find evidence from multiple independent photometric datasets that this star has a strong differential rotation signal, indicating that a singly (quasi)-periodic activity model is insufficient. This explains why more complex models were favored over simpler models in SM21, even though the preferred model fell victim to overfitting.
\end{itemize}

The first point, in particular, warrants further scrutiny for stars across a range of ages and spectral types. We argued in Section \ref{sec:why} that RV datasets taken by instruments with different bandpasses and calculated using different RV extraction techniques should be linear combinations of each other, recapitulating the observation made in \citet{Cale:2021aa}, but this assumption may not be true. Contemporaneous RV datasets made by different instruments will help test this assumption.

These authors have devoted significant person- and computer-power to producing a fit to the data presented here that take into account all of these effects. However, we have found that jointly fitting all the data using only a single rotation period forces all of the instrumental GP amplitudes to 0. We interpret this as evidence that a singly (quasi)-periodic GP model is incapable of fitting the data (i.e., a more complex model is needed), and differential rotation provides a ready (but not sole) explanation. However, the differential rotation effects are very complicated to disentangle with the current dataset.\footnote{Although we highly encourage others to try!} Again, we suggest a high-cadence RV campaign to resolve the multiple, nearby periodicities in the RVs and construct a high-fidelity model.

One important detail to note is that the GP kernel which best-fits a highly active, rapid-rotator like V1298 Tau may be wholly inappropriate to fit the activity signal of an older, quieter, Sun-like star. In young, rapid-rotators, the activity signal is relatively long-lived, often stable across several observation epochs \citep[e.g.,][]{Yu_V410tau_2019, Carvalho_hubble4_2021}. 

On the other hand, Sun-like stars have much shorter-lived spots, sometimes evolving over the course of one or two week observing campaigns \citep{Giles_KeplerLifetimes_2017MNRAS, Namekata_DecayRatesOfStarspots_2019ApJ}. A GP kernel describing the activity of Sun-like stars should be more flexible, allowing for more rapidly changing and decaying signals. While a single kernel may be capable of spanning these regimes of period evolution, the attempt to construct one should be made with caution. For the time being, the best approach may be to treat the two regimes of activity with unique kernels.

This analysis is imperfect and incomplete. Many of the effects we have discussed are subtle, and we encourage others to study them further. This analysis has also evolved (quite a lot) over the preparation of this study.

There are many exciting follow-up avenues for the V1298~Tau system. First, an independent determination of the planet masses with TTVs would be enormously helpful in providing a ``check'' for RV modelers. Second, we believe it is worthwhile to explore modeling frameworks for V1298 Tau that explicitly model the relationship between contemporaneous photometry, activity indices, and multiple RV datasets. These frameworks (such as that of \citealt{Rajpaul:2015aa} and \citealt{Cale:2021aa}) move beyond sharing hyperparameters between contemporaneous photometric and RV datasets and allow a function of one dataset to be directly correlated with the other, decreasing the overfitting potential. In the longer term, comparing or jointly modeling these data with Doppler tomographic information and spectrum-level measurements, as in \citet{ Yu_v410Tau_2019MNRAS, Finociety_V410Spirou_2021, Klein:2022aa} will provide even stronger constraints.

In addition to working toward an optimal physical model of all available data, it is worth investigating alternative statistical modeling pathways to GPR, especially low computational cost techniques like autoregressive moving average (ARMA) models (\citealt{Feigelson:2018a}, \citealt{Durbin:2001a}). ARMA models treat the $i$th datapoint as a linear combination of past data points and model residuals, and ``training'' involves optimizing the linear coefficients. Directly comparing models constructed with ARMA and GPR would be a worthwhile exercise in general for datasets containing stellar activity, and in particular for young, active stars.

We believe that understanding the RV variability of young stars is an endeavor that will pay dividends in the near future. The relative long-term stability of activity on young stars allows for detailed study of a given spot geometry and its impact on both photometric and spectroscopic observations across multiple bands. As we work to understand how to best fit activity with GPs, young stars, particularly WTTSs, provide good laboratories on which to test our techniques. 

Just as we validate the performance of a new instrument on stars with large, well-studied Keplerian signals, we must, as a field, validate the performance of our activity-modeling techniques on stars with large, well-studied activity signals before we can trust activity-models applied to Sun-like stars at 30 \cms{} precision\footnote{In fact, the activity-to-Keplerian ratio of 1000 \ms{}: 50 \ms{} for warm giant planets around a young star like V1298~Tau is reminiscent of the 1 \ms{} : 10 \cms{} ratio for an Earth around a Sun-like star.}. This starts by allocating resources to the construction of high-cadence RV datasets of young stars, and continues by studying a) the relationship between RVs and auxiliary data, such as photometry and activity indices, b) the best phenomenological models (kernels, etc) for the data, c) the best methods for validating a given model's accuracy, and d) the cadence needed to resolve periodic signals (and combinations of signals). We believe that these studies, on young stars, will pave the way for stellar activity models with 30 \cms{} predictive capability, on which the characterization of Earth 2.0 depends.

\appendix

\begin{figure*}
    \centering
    \hspace*{-0.5in}
    \vspace*{-5.5in}
    \includegraphics[trim={0 0 0 0},clip]{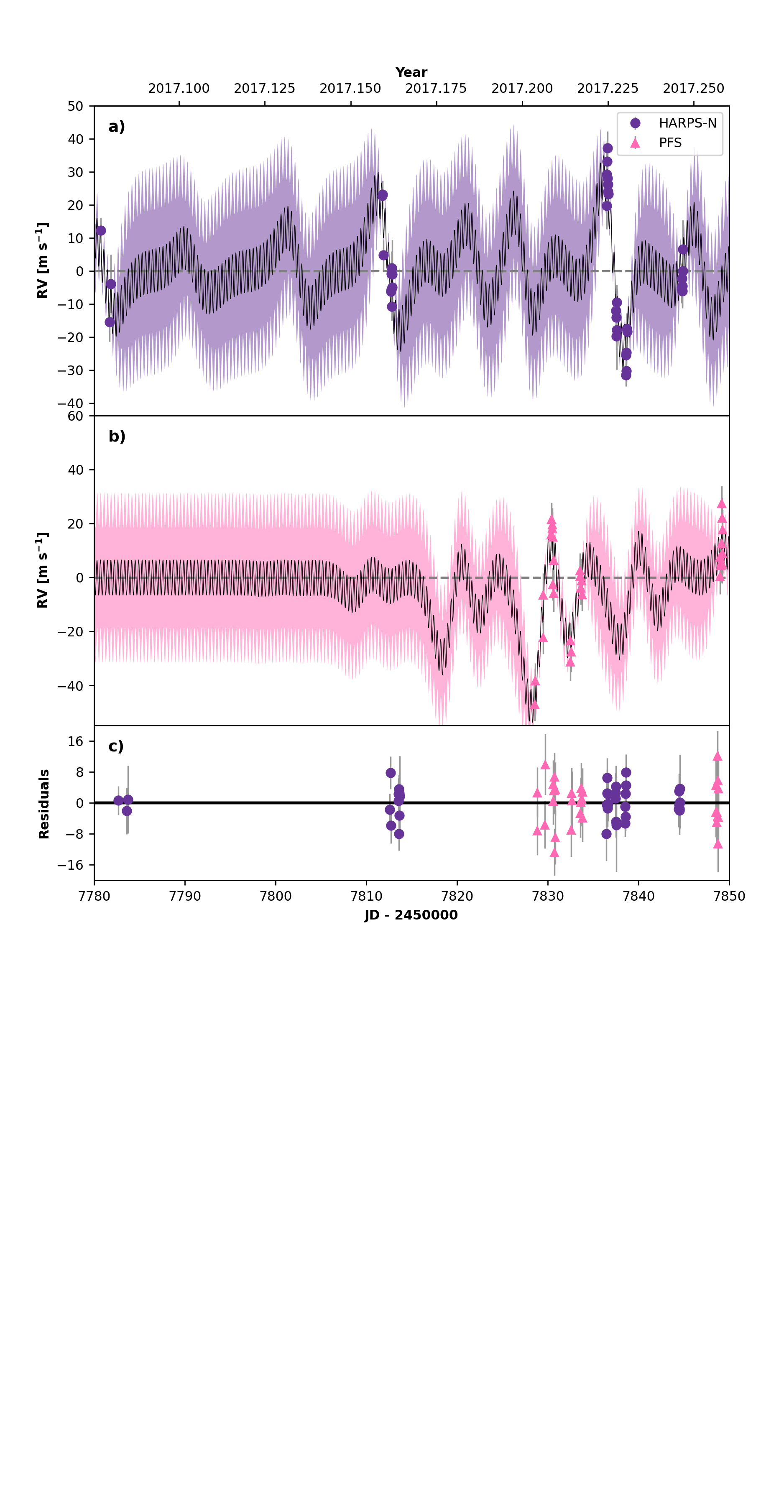}
    \caption{Demonstration of the impact of constructing separate covariance matrices and adding the log(likelihoods). Compare with Figure \ref{fig:radvel_mult}. The data and best-fit parameters are for K2-131, published in \citet{Dai:2017aa}, for demonstration purposes only. Top: GP mean prediction (black solid line) and 1-$\sigma$ uncertainties (purple filled), together with the HARPS-N data points on which the GP is conditioned (purple points). Middle: Same as top, but for PFS data. Bottom: Residuals with respect to the GP mean prediction. Takeaway: When separate covariance matrices for each RV instrument are used, contemporaneous data are uncorrelated in the model, allowing additional degrees of freedom.}
    \label{fig:radvel_indep}
\end{figure*}

\begin{figure*}
    \centering
    \hspace*{-0.5in}
    \vspace*{-5.5in}
    \includegraphics[trim={0 0 0 0},clip]{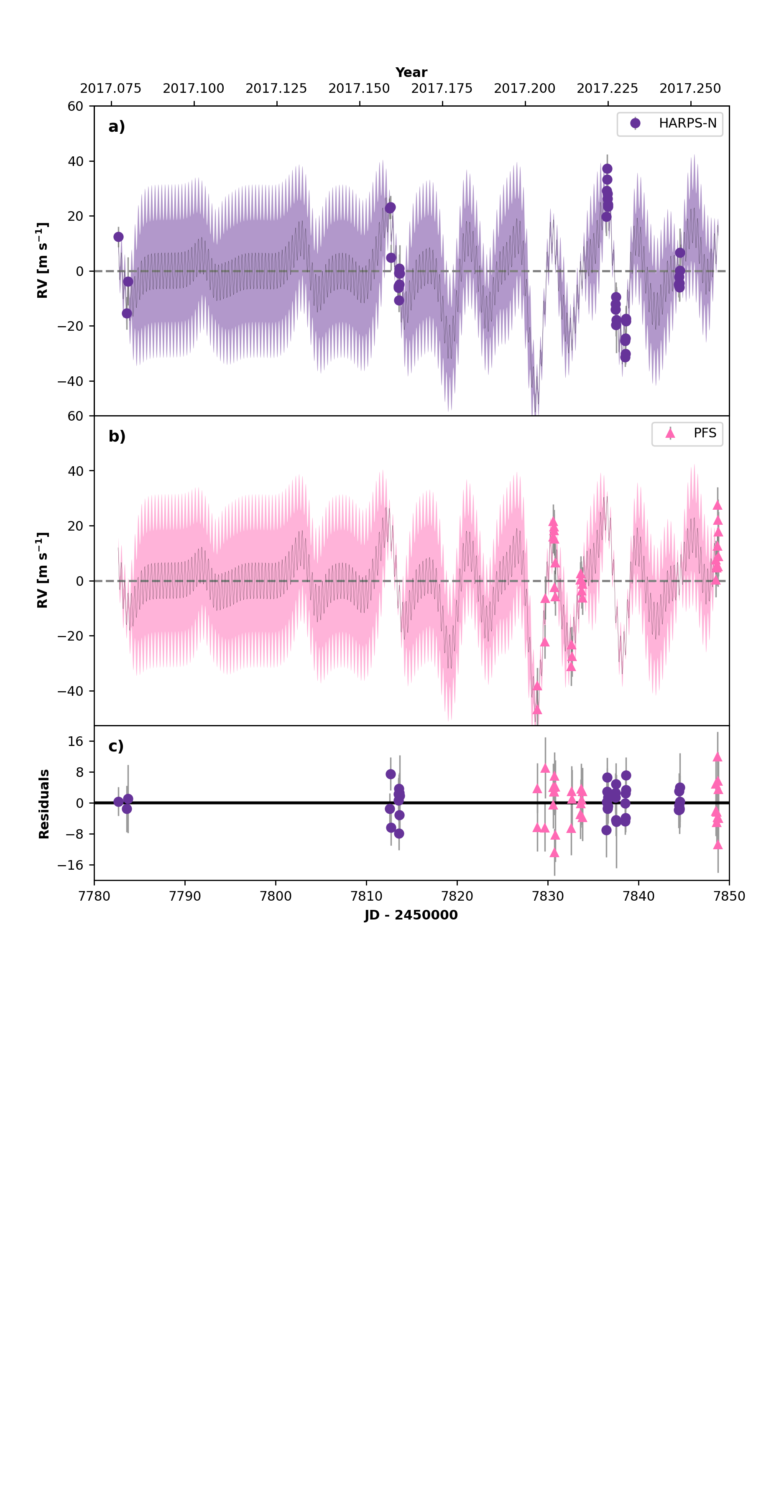}
    \caption{Same as Figure \ref{fig:radvel_indep} (in particular, using the exact same data and GP hyperparameters), but here a single covariance matrix is constructed, following the suggestion in Section \ref{sec:corr_data}. Takeaway: Constructing a single covariance matrix requires that GP predictions for separate instruments are scalar multiples of one another, which is more consistent with physical expectations and results in a more constrained model than one with a separate covariance matrix for each instrument.}
    \label{fig:radvel_mult}
\end{figure*}

\section{Gaussian Processes and Occam's Razor}

Many introductions to GPR (e.g., \citealt{Aigrain:2022aa}) mention that the GP likelihood has an ``Occam's razor'' term built in that penalizes complexity. This section briefly reviews GPR, then outlines a geometrical interpretation of the complexity penalty in order to further readers' understanding.

A Gaussian process regression model parameterizes the covariance between data points using a kernel function. A statistician may pick an arbitrary function (subject to certain mathematical requirements, see \citealt{Rasmussen:2006aa} for the gory details) to be the kernel, which can then be used to calculate the covariance between any two data points. As an example, let's consider the periodic kernel:

\begin{equation}
\label{eq:per}
    C_{ij} = \eta_1^2 \exp\left[-
                 \frac{ \sin^2(\frac{ \pi|t_i-t_j| }{ P_{\rm rot} } ) }{ \eta_3^2 } \right],
\end{equation}where $\eta_1$ is the amplitude, $P_{\rm rot}$ is the variability period (often the star's rotation period), and $\eta_3$ is the harmonic complexity, or degree of ``wiggly-ness'' of the repeating signal. Given this model for the covariance of our data, and some data, we can make a prediction, which is the conditional probability distribution over expected values at new measurement times. This is referred to as \textit{conditioning} a GP on a set of data. 

Importantly, Gaussian process regression does not inherently involve training (i.e., parameter tuning, generally via an optimization and/or MCMC step). Gaussian process regression is just the process of using a parametrization of your covariance matrix to predict the values and uncertainties of new data points given existing data points. 

The ``training'' part comes in when you are optimizing the hyperparameters of your kernel (optionally jointly with parameters of a mean function, which could be a function of Keplerian orbital parameters). Now, it becomes important to compute a statistic describing how well your GP model fits your data, so that you can optimize the (hyper)parameters to obtain your result. This is where the Gaussian process likelihood comes in:

\begin{equation}
\label{eq:likelihood}
    \log{p(d|m)} = -\frac{N}{2}\log{2\pi} -  \frac{1}{2} \textbf{r
    }^T \textbf{C}^{-1} \textbf{r} - \frac{1}{2}\log{|\textbf{C}|},
\end{equation}where \textbf{C} is the covariance matrix computed for the times at which you have data, $N$ is the number of measurements, and \textbf{r} is the vector of residuals (data - mean model). The first term is a constant, and does not change as a function of the kernel hyperparameters, and the second term is analogous to $\chi^2$ (in fact, it reduces to $\chi^2$ in the limit of no off-diagonal covariance). The second term describes how well your mean model and correlated noise description matches your data. The third term is the ``Occam's razor'' term that penalizes complexity. 

To understand how the third term penalizes complexity, recall that the determinant of a matrix can be understood as the hypervolume between vectors defined by the columns of the matrix. To make this concrete, consider the 3x3 identity matrix:

\begin{equation}
    \begin{pmatrix}
        1 & 0 & 0 \\
        0 & 1 & 0 \\
        0 & 0 & 1 \\
    \end{pmatrix}.
\end{equation} The vectors defined by the columns of this matrix are (1,0,0), (0,1,0), and (0,0,1). The volume of the 3D shape defined by these vectors (the unit cube) is 1, the same as the matrix determinant!

The \textit{i}-th column vector of a covariance matrix can be interpreted as the vector of covariances between a data point taken at t$_i$ and every other data point in the dataset. The determinant of this matrix, then, is the hypervolume defined by these covariance vectors. A perfectly covariant matrix, in which all data points are perfectly correlated, will consist of all 1s\footnote{Or a scalar multiple of the matrix of all 1s.}, and the covariance vectors will all ``point'' in the same direction. This results in a third-term contribution of:

\begin{equation}
\begin{split}
    - \frac{1}{2}\log{|\textbf{C}|} & = - \frac{1}{2}\log{0} \\ & = - (-\infty) \\ & = \infty.
\end{split}
\end{equation} A matrix of perfectly \textit{independent} data points, on the other hand, is (a scalar multiple of) the identity matrix. The covariance vectors all ``point'' in orthogonal directions. This matrix results in a third-term contribution of:

\begin{equation}
\begin{split}
    - \frac{1}{2}\log{|\textbf{C}|} & = - \frac{1}{2}\log{1} \\ & = 0.
\end{split}
\end{equation}This exercise demonstrates that the determinant of the covariance matrix quantifies how ``clustered'' the covariance vectors corresponding to each data point are in hyperspace. More clustered covariance vectors get a big likelihood boost, while less clustered/more independent covariance vectors get a smaller boost. Figure 5.3 in \cite{Rasmussen:2006aa} decomposes the likelihood contributions of the second and third terms in Equation \ref{eq:likelihood}, illustrating how they combine to produce a local likelihood maximum in parameter space for a toy model. 

\begin{acknowledgements}

S.B. wishes to thank first and foremost Alejandro Su\'{a}rez-Mascare\~{n}o for constructive and helpful thoughts throughout the preparation of this study. S.B. also wishes to thank the small army of people who shaped this analysis through conversation: Jason Wang and his research group, Heather Knutson's research group, the folks at the Flatiron CCA, the University of Michigan Exoplanet Journal Club, the UC Riverside Astrobiology Seminar group, Johanna Teske and the astronomers of the Carnegie Earth \& Planets Laboratory, J\'{e}a Adams, Kim Paragas, Shreyas Vissapragada, Ward Howard, and Roberto Tejada Arevalo. All of the authors thank both the anonymous referee and the anonymous statistics editor for helpful comments that made us further question our assumptions and improved this work.

J.M.A.M. is supported by the National Science Foundation Graduate Research Fellowship Program under Grant No. DGE-1842400. J.M.A.M. acknowledges the LSSTC Data Science Fellowship Program, which is funded by LSSTC, NSF Cybertraining Grant No. 1829740, the Brinson Foundation, and the Moore Foundation; his participation in the program has benefited this work. T.H. is supported by JSPS KAKENHI Grant Numbers JP19K14783 and JP21H00035.

This research was enabled by the following software: \texttt{numpy} \citep{harris2020array}, \texttt{Lightkurve}, a Python package for Kepler and TESS data analysis \citep{lightkurve:2018aa}, \texttt{pandas} \citep{pandas}, \texttt{matplotlib} \citep{Hunter:2007aa}, \texttt{scipy} \citep{scipy}, \texttt{astropy} (\citealt{astropy:2013}, \citealt{astropy:2018}, \citealt{astropy:2022}), \texttt{jax} \citep{jax2018github}, \texttt{george} \citep{hodlr}, \texttt{celerite} \citep{celerite}, \texttt{tinygp} (github.com/dfm/tinygp), and \texttt{radvel} \citep{Fulton:2018aa}.

S.B. wishes to acknowledge her status as a settler on the ancestral lands of the Gabrielino/Tongva people, and to recognize that the astronomical observations described in this paper were only possible because of the dispossession of Maunakea from Kan\={a}ka Maoli. We seek to work toward a scientific practice guided by pono and a future in which we all honor the land.

\end{acknowledgements}

\bibliographystyle{aasjournal}
\bibliography{V1298Tau}

\begin{thebibliography}{}
\expandafter\ifx\csname natexlab\endcsname\relax\def\natexlab#1{#1}\fi
\providecommand{\url}[1]{\href{#1}{#1}}
\providecommand{\dodoi}[1]{doi:~\href{http://doi.org/#1}{\nolinkurl{#1}}}
\providecommand{\doeprint}[1]{\href{http://ascl.net/#1}{\nolinkurl{http://ascl.net/#1}}}
\providecommand{\doarXiv}[1]{\href{https://arxiv.org/abs/#1}{\nolinkurl{https://arxiv.org/abs/#1}}}

\bibitem[{{Aigrain} \& {Foreman-Mackey}(2022)}]{Aigrain:2022aa}
{Aigrain}, S., \& {Foreman-Mackey}, D. 2022, arXiv e-prints, arXiv:2209.08940.
\newblock \doarXiv{2209.08940}

\bibitem[{{Aigrain} {et~al.}(2012){Aigrain}, {Pont}, \&
  {Zucker}}]{Aigrain:2012aa}
{Aigrain}, S., {Pont}, F., \& {Zucker}, S. 2012, \mnras, 419, 3147,
  \dodoi{10.1111/j.1365-2966.2011.19960.x}

\bibitem[{{Ambikasaran} {et~al.}(2014){Ambikasaran}, {Foreman-Mackey},
  {Greengard}, {Hogg}, \& {O'Neil}}]{hodlr}
{Ambikasaran}, S., {Foreman-Mackey}, D., {Greengard}, L., {Hogg}, D.~W., \&
  {O'Neil}, M. 2014

\bibitem[{{Astropy Collaboration} {et~al.}(2013){Astropy Collaboration},
  {Robitaille}, {Tollerud}, {Greenfield}, {Droettboom}, {Bray}, {Aldcroft},
  {Davis}, {Ginsburg}, {Price-Whelan}, {Kerzendorf}, {Conley}, {Crighton},
  {Barbary}, {Muna}, {Ferguson}, {Grollier}, {Parikh}, {Nair}, {Unther},
  {Deil}, {Woillez}, {Conseil}, {Kramer}, {Turner}, {Singer}, {Fox}, {Weaver},
  {Zabalza}, {Edwards}, {Azalee Bostroem}, {Burke}, {Casey}, {Crawford},
  {Dencheva}, {Ely}, {Jenness}, {Labrie}, {Lim}, {Pierfederici}, {Pontzen},
  {Ptak}, {Refsdal}, {Servillat}, \& {Streicher}}]{astropy:2013}
{Astropy Collaboration}, {Robitaille}, T.~P., {Tollerud}, E.~J., {et~al.} 2013,
  \aap, 558, A33, \dodoi{10.1051/0004-6361/201322068}

\bibitem[{{Astropy Collaboration} {et~al.}(2018){Astropy Collaboration},
  {Price-Whelan}, {Sip{\H{o}}cz}, {G{\"u}nther}, {Lim}, {Crawford}, {Conseil},
  {Shupe}, {Craig}, {Dencheva}, {Ginsburg}, {Vand erPlas}, {Bradley},
  {P{\'e}rez-Su{\'a}rez}, {de Val-Borro}, {Aldcroft}, {Cruz}, {Robitaille},
  {Tollerud}, {Ardelean}, {Babej}, {Bach}, {Bachetti}, {Bakanov}, {Bamford},
  {Barentsen}, {Barmby}, {Baumbach}, {Berry}, {Biscani}, {Boquien}, {Bostroem},
  {Bouma}, {Brammer}, {Bray}, {Breytenbach}, {Buddelmeijer}, {Burke},
  {Calderone}, {Cano Rodr{\'\i}guez}, {Cara}, {Cardoso}, {Cheedella}, {Copin},
  {Corrales}, {Crichton}, {D'Avella}, {Deil}, {Depagne}, {Dietrich}, {Donath},
  {Droettboom}, {Earl}, {Erben}, {Fabbro}, {Ferreira}, {Finethy}, {Fox},
  {Garrison}, {Gibbons}, {Goldstein}, {Gommers}, {Greco}, {Greenfield},
  {Groener}, {Grollier}, {Hagen}, {Hirst}, {Homeier}, {Horton}, {Hosseinzadeh},
  {Hu}, {Hunkeler}, {Ivezi{\'c}}, {Jain}, {Jenness}, {Kanarek}, {Kendrew},
  {Kern}, {Kerzendorf}, {Khvalko}, {King}, {Kirkby}, {Kulkarni}, {Kumar},
  {Lee}, {Lenz}, {Littlefair}, {Ma}, {Macleod}, {Mastropietro}, {McCully},
  {Montagnac}, {Morris}, {Mueller}, {Mumford}, {Muna}, {Murphy}, {Nelson},
  {Nguyen}, {Ninan}, {N{\"o}the}, {Ogaz}, {Oh}, {Parejko}, {Parley}, {Pascual},
  {Patil}, {Patil}, {Plunkett}, {Prochaska}, {Rastogi}, {Reddy Janga},
  {Sabater}, {Sakurikar}, {Seifert}, {Sherbert}, {Sherwood-Taylor}, {Shih},
  {Sick}, {Silbiger}, {Singanamalla}, {Singer}, {Sladen}, {Sooley},
  {Sornarajah}, {Streicher}, {Teuben}, {Thomas}, {Tremblay}, {Turner},
  {Terr{\'o}n}, {van Kerkwijk}, {de la Vega}, {Watkins}, {Weaver}, {Whitmore},
  {Woillez}, {Zabalza}, \& {Astropy Contributors}}]{astropy:2018}
{Astropy Collaboration}, {Price-Whelan}, A.~M., {Sip{\H{o}}cz}, B.~M., {et~al.}
  2018, \aj, 156, 123, \dodoi{10.3847/1538-3881/aabc4f}

\bibitem[{{Astropy Collaboration} {et~al.}(2022){Astropy Collaboration},
  {Price-Whelan}, {Lim}, {Earl}, {Starkman}, {Bradley}, {Shupe}, {Patil},
  {Corrales}, {Brasseur}, {N{"o}the}, {Donath}, {Tollerud}, {Morris},
  {Ginsburg}, {Vaher}, {Weaver}, {Tocknell}, {Jamieson}, {van Kerkwijk},
  {Robitaille}, {Merry}, {Bachetti}, {G{"u}nther}, {Aldcroft},
  {Alvarado-Montes}, {Archibald}, {B{'o}di}, {Bapat}, {Barentsen}, {Baz{'a}n},
  {Biswas}, {Boquien}, {Burke}, {Cara}, {Cara}, {Conroy}, {Conseil}, {Craig},
  {Cross}, {Cruz}, {D'Eugenio}, {Dencheva}, {Devillepoix}, {Dietrich},
  {Eigenbrot}, {Erben}, {Ferreira}, {Foreman-Mackey}, {Fox}, {Freij}, {Garg},
  {Geda}, {Glattly}, {Gondhalekar}, {Gordon}, {Grant}, {Greenfield}, {Groener},
  {Guest}, {Gurovich}, {Handberg}, {Hart}, {Hatfield-Dodds}, {Homeier},
  {Hosseinzadeh}, {Jenness}, {Jones}, {Joseph}, {Kalmbach}, {Karamehmetoglu},
  {Ka{l}uszy{'n}ski}, {Kelley}, {Kern}, {Kerzendorf}, {Koch}, {Kulumani},
  {Lee}, {Ly}, {Ma}, {MacBride}, {Maljaars}, {Muna}, {Murphy}, {Norman},
  {O'Steen}, {Oman}, {Pacifici}, {Pascual}, {Pascual-Granado}, {Patil},
  {Perren}, {Pickering}, {Rastogi}, {Roulston}, {Ryan}, {Rykoff}, {Sabater},
  {Sakurikar}, {Salgado}, {Sanghi}, {Saunders}, {Savchenko}, {Schwardt},
  {Seifert-Eckert}, {Shih}, {Jain}, {Shukla}, {Sick}, {Simpson},
  {Singanamalla}, {Singer}, {Singhal}, {Sinha}, {Sip{H{o}}cz}, {Spitler},
  {Stansby}, {Streicher}, {{{S}}umak}, {Swinbank}, {Taranu}, {Tewary},
  {Tremblay}, {Val-Borro}, {Van Kooten}, {Vasovi{'c}}, {Verma}, {de Miranda
  Cardoso}, {Williams}, {Wilson}, {Winkel}, {Wood-Vasey}, {Xue}, {Yoachim},
  {Zhang}, {Zonca}, \& {Astropy Project Contributors}}]{astropy:2022}
{Astropy Collaboration}, {Price-Whelan}, A.~M., {Lim}, P.~L., {et~al.} 2022,
  apj, 935, 167, \dodoi{10.3847/1538-4357/ac7c74}

\bibitem[{{Basri} \& {Shah}(2020)}]{Basri:2020aa}
{Basri}, G., \& {Shah}, R. 2020, \apj, 901, 14,
  \dodoi{10.3847/1538-4357/abae5d}

\bibitem[{{Benatti} {et~al.}(2019){Benatti}, {Nardiello}, {Malavolta},
  {Desidera}, {Borsato}, {Nascimbeni}, {Damasso}, {D'Orazi}, {Mesa}, {Messina},
  {Esposito}, {Bignamini}, {Claudi}, {Covino}, {Lovis}, \&
  {Sabotta}}]{Benatti:2019aa}
{Benatti}, S., {Nardiello}, D., {Malavolta}, L., {et~al.} 2019, \aap, 630, A81,
  \dodoi{10.1051/0004-6361/201935598}

\bibitem[{{Bouma} {et~al.}(2022){Bouma}, {Curtis}, {Masuda}, {Hillenbrand},
  {Stefansson}, {Isaacson}, {Narita}, {Fukui}, {Ikoma}, {Tamura}, {Kraus},
  {Furlan}, {Gnilka}, {Lester}, \& {Howell}}]{Bouma:2022aa}
{Bouma}, L.~G., {Curtis}, J.~L., {Masuda}, K., {et~al.} 2022, \aj, 163, 121,
  \dodoi{10.3847/1538-3881/ac4966}

\bibitem[{Bradbury {et~al.}(2018)Bradbury, Frostig, Hawkins, Johnson, Leary,
  Maclaurin, Necula, Paszke, Vander{P}las, Wanderman-{M}ilne, \&
  Zhang}]{jax2018github}
Bradbury, J., Frostig, R., Hawkins, P., {et~al.} 2018, {JAX}: composable
  transformations of {P}ython+{N}um{P}y programs, 0.3.13.
\newblock \url{http://github.com/google/jax}

\bibitem[{{Burke} {et~al.}(2020){Burke}, {Levine}, {Fausnaugh}, {Vanderspek},
  {Barclay}, {Libby-Roberts}, {Morris}, {Sipocz}, {Owens}, {Feinstein}, \&
  {Camacho}}]{Burke:2020aa}
{Burke}, C.~J., {Levine}, A., {Fausnaugh}, M., {et~al.} 2020, {TESS-Point: High
  precision TESS pointing tool}, Astrophysics Source Code Library, record
  ascl:2003.001.
\newblock \doeprint{2003.001}

\bibitem[{{Butler} {et~al.}(1996){Butler}, {Marcy}, {Williams}, {McCarthy},
  {Dosanjh}, \& {Vogt}}]{Butler:1996aa}
{Butler}, R.~P., {Marcy}, G.~W., {Williams}, E., {et~al.} 1996, \pasp, 108,
  500, \dodoi{10.1086/133755}

\bibitem[{{Caceres} {et~al.}(2019){Caceres}, {Feigelson}, {Jogesh Babu},
  {Bahamonde}, {Christen}, {Bertin}, {Meza}, \& {Cur{\'e}}}]{Caceres:2019a}
{Caceres}, G.~A., {Feigelson}, E.~D., {Jogesh Babu}, G., {et~al.} 2019, \aj,
  158, 58, \dodoi{10.3847/1538-3881/ab26ba}

\bibitem[{{Cale} {et~al.}(2021){Cale}, {Reefe}, {Plavchan}, {Tanner}, {Gaidos},
  {Gagn{\'e}}, {Gao}, {Kane}, {B{\'e}jar}, {Lodieu}, {Anglada-Escud{\'e}},
  {Ribas}, {Pall{\'e}}, {Quirrenbach}, {Amado}, {Reiners}, {Caballero}, {Rosa
  Zapatero Osorio}, {Dreizler}, {Howard}, {Fulton}, {Xuesong Wang}, {Collins},
  {El Mufti}, {Wittrock}, {Gilbert}, {Barclay}, {Klein}, {Martioli},
  {Wittenmyer}, {Wright}, {Addison}, {Hirano}, {Tamura}, {Kotani}, {Narita},
  {Vermilion}, {Lee}, {Geneser}, {Teske}, {Quinn}, {Latham}, {Esquerdo},
  {Calkins}, {Berlind}, {Zohrabi}, {Stibbards}, {Kotnana}, {Jenkins},
  {Twicken}, {Henze}, {Kidwell}, {Burke}, {Villase{\~n}or}, \&
  {Boyd}}]{Cale:2021aa}
{Cale}, B.~L., {Reefe}, M., {Plavchan}, P., {et~al.} 2021, \aj, 162, 295,
  \dodoi{10.3847/1538-3881/ac2c80}

\bibitem[{{Carpenter} {et~al.}(2001){Carpenter}, {Hillenbrand}, \&
  {Skrutskie}}]{Carpenter_NIRPhotometricVar_2001AJ}
{Carpenter}, J.~M., {Hillenbrand}, L.~A., \& {Skrutskie}, M.~F. 2001, \aj, 121,
  3160, \dodoi{10.1086/32108610.48550/arXiv.astro-ph/0102446}

\bibitem[{{Carvalho} {et~al.}(2021){Carvalho}, {Johns-Krull}, {Prato}, \&
  {Anderson}}]{Carvalho_hubble4_2021}
{Carvalho}, A., {Johns-Krull}, C.~M., {Prato}, L., \& {Anderson}, J. 2021,
  \apj, 910, 33, \dodoi{10.3847/1538-4357/abe237}

\bibitem[{{Chaplin} \& {Miglio}(2013)}]{Chaplin2013aa}
{Chaplin}, W.~J., \& {Miglio}, A. 2013, \araa, 51, 353,
  \dodoi{10.1146/annurev-astro-082812-140938}

\bibitem[{{Chen} \& {Rogers}(2016)}]{Chen:2016aa}
{Chen}, H., \& {Rogers}, L.~A. 2016, \apj, 831, 180,
  \dodoi{10.3847/0004-637X/831/2/180}

\bibitem[{{Dai} {et~al.}(2017){Dai}, {Winn}, {Gandolfi}, {Wang}, {Teske},
  {Burt}, {Albrecht}, {Barrag{\'a}n}, {Cochran}, {Endl}, {Fridlund}, {Hatzes},
  {Hirano}, {Hirsch}, {Johnson}, {Justesen}, {Livingston}, {Persson},
  {Prieto-Arranz}, {Vanderburg}, {Alonso}, {Antoniciello}, {Arriagada},
  {Butler}, {Cabrera}, {Crane}, {Cusano}, {Csizmadia}, {Deeg}, {Dieterich},
  {Eigm{\"u}ller}, {Erikson}, {Everett}, {Fukui}, {Grziwa}, {Guenther},
  {Henry}, {Howell}, {Johnson}, {Korth}, {Kuzuhara}, {Narita}, {Nespral},
  {Nowak}, {Palle}, {P{\"a}tzold}, {Rauer}, {Monta{\~n}{\'e}s Rodr{\'\i}guez},
  {Shectman}, {Smith}, {Thompson}, {Van Eylen}, {Williamson}, \&
  {Wittenmyer}}]{Dai:2017aa}
{Dai}, F., {Winn}, J.~N., {Gandolfi}, D., {et~al.} 2017, \aj, 154, 226,
  \dodoi{10.3847/1538-3881/aa9065}

\bibitem[{{David} {et~al.}(2019{\natexlab{a}}){David}, {Petigura}, {Luger},
  {Foreman-Mackey}, {Livingston}, {Mamajek}, \& {Hillenbrand}}]{David2019ab}
{David}, T.~J., {Petigura}, E.~A., {Luger}, R., {et~al.} 2019{\natexlab{a}},
  \apjl, 885, L12, \dodoi{10.3847/2041-8213/ab4c99}

\bibitem[{{David} {et~al.}(2016){David}, {Hillenbrand}, {Petigura},
  {Carpenter}, {Crossfield}, {Hinkley}, {Ciardi}, {Howard}, {Isaacson}, {Cody},
  {Schlieder}, {Beichman}, \& {Barenfeld}}]{David2016aa}
{David}, T.~J., {Hillenbrand}, L.~A., {Petigura}, E.~A., {et~al.} 2016, \nat,
  534, 658, \dodoi{10.1038/nature18293}

\bibitem[{{David} {et~al.}(2019{\natexlab{b}}){David}, {Cody}, {Hedges},
  {Mamajek}, {Hillenbrand}, {Ciardi}, {Beichman}, {Petigura}, {Fulton},
  {Isaacson}, {Howard}, {Gagn{\'e}}, {Saunders}, {Rebull}, {Stauffer},
  {Vasisht}, \& {Hinkley}}]{David2019aa}
{David}, T.~J., {Cody}, A.~M., {Hedges}, C.~L., {et~al.} 2019{\natexlab{b}},
  \aj, 158, 79, \dodoi{10.3847/1538-3881/ab290f}

\bibitem[{{Durbin} \& {Koopman}(2001)}]{Durbin:2001a}
{Durbin}, J., \& {Koopman}, S.~J. 2001, {Time Series Analysis by State Space
  Methods}

\bibitem[{{Espinoza} {et~al.}(2018){Espinoza}, {Kossakowski}, \&
  {Brahm}}]{juliet}
{Espinoza}, N., {Kossakowski}, D., \& {Brahm}, R. 2018, arXiv e-prints,
  arXiv:1812.08549.
\newblock \doarXiv{1812.08549}

\bibitem[{{Fabrycky} {et~al.}(2014){Fabrycky}, {Lissauer}, {Ragozzine}, {Rowe},
  {Steffen}, {Agol}, {Barclay}, {Batalha}, {Borucki}, {Ciardi}, {Ford},
  {Gautier}, {Geary}, {Holman}, {Jenkins}, {Li}, {Morehead}, {Morris},
  {Shporer}, {Smith}, {Still}, \& {Van Cleve}}]{Fabrycky:2014aa}
{Fabrycky}, D.~C., {Lissauer}, J.~J., {Ragozzine}, D., {et~al.} 2014, \apj,
  790, 146, \dodoi{10.1088/0004-637X/790/2/14610.48550/arXiv.1202.6328}

\bibitem[{{Feigelson} {et~al.}(2018){Feigelson}, {Babu}, \&
  {Caceres}}]{Feigelson:2018a}
{Feigelson}, E.~D., {Babu}, G.~J., \& {Caceres}, G.~A. 2018, Frontiers in
  Physics, 6, 80, \dodoi{10.3389/fphy.2018.00080}

\bibitem[{{Feinstein} {et~al.}(2022){Feinstein}, {David}, {Montet},
  {Foreman-Mackey}, {Livingston}, \& {Mann}}]{Feinstein2022aa}
{Feinstein}, A.~D., {David}, T.~J., {Montet}, B.~T., {et~al.} 2022, \apjl, 925,
  L2, \dodoi{10.3847/2041-8213/ac4745}

\bibitem[{{Finociety} {et~al.}(2021){Finociety}, {Donati}, {Klein}, {Zaire},
  {Lehmann}, {Moutou}, {Bouvier}, {Alencar}, {Yu}, {Grankin}, {Artigau},
  {Doyon}, {Delfosse}, {Fouqu{\'e}}, {H{\'e}brard}, {Jardine},
  {K{\'o}sp{\'a}l}, {M{\'e}nard}, {M{\'e}nard}, \& {SLS
  Consortium}}]{Finociety_V410Spirou_2021}
{Finociety}, B., {Donati}, J.~F., {Klein}, B., {et~al.} 2021, \mnras, 508,
  3427, \dodoi{10.1093/mnras/stab2778}

\bibitem[{{Foreman-Mackey} {et~al.}(2017){Foreman-Mackey}, {Agol}, {Angus}, \&
  {Ambikasaran}}]{celerite}
{Foreman-Mackey}, D., {Agol}, E., {Angus}, R., \& {Ambikasaran}, S. 2017,
  ArXiv.
\newblock \url{https://arxiv.org/abs/1703.09710}

\bibitem[{Foreman-Mackey {et~al.}(2022)Foreman-Mackey, Yadav, theorashid,
  Fowlie, Tronsgaard, Schmerler, \&
  Killestein}]{dan_foreman_mackey_2022_7269074}
Foreman-Mackey, D., Yadav, S., theorashid, {et~al.} 2022, dfm/tinygp: v0.2.3,
  v0.2.3,  Zenodo, \dodoi{10.5281/zenodo.7269074}

\bibitem[{{Frasca} {et~al.}(2011){Frasca}, {Fr{\"o}hlich}, {Bonanno},
  {Catanzaro}, {Biazzo}, \& {Molenda-{\.Z}akowicz}}]{Frasca2011aa}
{Frasca}, A., {Fr{\"o}hlich}, H.~E., {Bonanno}, A., {et~al.} 2011, \aap, 532,
  A81, \dodoi{10.1051/0004-6361/201116980}

\bibitem[{{Fulton} {et~al.}(2018){Fulton}, {Petigura}, {Blunt}, \&
  {Sinukoff}}]{Fulton:2018aa}
{Fulton}, B.~J., {Petigura}, E.~A., {Blunt}, S., \& {Sinukoff}, E. 2018, \pasp,
  130, 044504, \dodoi{10.1088/1538-3873/aaaaa8}

\bibitem[{{Fulton} {et~al.}(2016){Fulton}, {Howard}, {Weiss}, {Sinukoff},
  {Petigura}, {Isaacson}, {Hirsch}, {Marcy}, {Henry}, {Grunblatt}, {Huber},
  {von Braun}, {Boyajian}, {Kane}, {Wittrock}, {Horch}, {Ciardi}, {Howell},
  {Wright}, \& {Ford}}]{Fulton:2016aa}
{Fulton}, B.~J., {Howard}, A.~W., {Weiss}, L.~M., {et~al.} 2016, \apj, 830, 46,
  \dodoi{10.3847/0004-637X/830/1/46}

\bibitem[{{Giles} {et~al.}(2017){Giles}, {Collier Cameron}, \&
  {Haywood}}]{Giles_KeplerLifetimes_2017MNRAS}
{Giles}, H. A.~C., {Collier Cameron}, A., \& {Haywood}, R.~D. 2017, \mnras,
  472, 1618, \dodoi{10.1093/mnras/stx1931}

\bibitem[{{Ginzburg} {et~al.}(2018){Ginzburg}, {Schlichting}, \&
  {Sari}}]{Ginzburg2018aa}
{Ginzburg}, S., {Schlichting}, H.~E., \& {Sari}, R. 2018, \mnras, 476, 759,
  \dodoi{10.1093/mnras/sty290}

\bibitem[{{Grunblatt} {et~al.}(2015){Grunblatt}, {Howard}, \&
  {Haywood}}]{Grunblatt:2015aa}
{Grunblatt}, S.~K., {Howard}, A.~W., \& {Haywood}, R.~D. 2015, \apj, 808, 127,
  \dodoi{10.1088/0004-637X/808/2/127}

\bibitem[{Harris {et~al.}(2020)Harris, Millman, van~der Walt, Gommers,
  Virtanen, Cournapeau, Wieser, Taylor, Berg, Smith, Kern, Picus, Hoyer, van
  Kerkwijk, Brett, Haldane, del R{\'{i}}o, Wiebe, Peterson,
  G{\'{e}}rard-Marchant, Sheppard, Reddy, Weckesser, Abbasi, Gohlke, \&
  Oliphant}]{harris2020array}
Harris, C.~R., Millman, K.~J., van~der Walt, S.~J., {et~al.} 2020, Nature, 585,
  357, \dodoi{10.1038/s41586-020-2649-2}

\bibitem[{{Howard} {et~al.}(2010){Howard}, {Marcy}, {Johnson}, {Fischer},
  {Wright}, {Isaacson}, {Valenti}, {Anderson}, {Lin}, \& {Ida}}]{Howard:2010aa}
{Howard}, A.~W., {Marcy}, G.~W., {Johnson}, J.~A., {et~al.} 2010, Science, 330,
  653, \dodoi{10.1126/science.1194854}

\bibitem[{Huerta {et~al.}(2008)Huerta, Johns-Krull, Prato, Hartigan, \&
  Jaffe}]{huerta_starspot-induced_2008}
Huerta, M., Johns-Krull, C.~M., Prato, L., Hartigan, P., \& Jaffe, D.~T. 2008,
  The Astrophysical Journal, 678, 472, \dodoi{10.1086/526415}

\bibitem[{Hunter(2007)}]{Hunter:2007aa}
Hunter, J.~D. 2007, Computing in Science \& Engineering, 9, 90,
  \dodoi{10.1109/MCSE.2007.55}

\bibitem[{{Jenkins} {et~al.}(2016){Jenkins}, {Twicken}, {McCauliff},
  {Campbell}, {Sanderfer}, {Lung}, {Mansouri-Samani}, {Girouard}, {Tenenbaum},
  {Klaus}, {Smith}, {Caldwell}, {Chacon}, {Henze}, {Heiges}, {Latham},
  {Morgan}, {Swade}, {Rinehart}, \& {Vanderspek}}]{Jenkins:2016aa}
{Jenkins}, J.~M., {Twicken}, J.~D., {McCauliff}, S., {et~al.} 2016, in Society
  of Photo-Optical Instrumentation Engineers (SPIE) Conference Series, Vol.
  9913, Software and Cyberinfrastructure for Astronomy IV, ed. G.~{Chiozzi} \&
  J.~C. {Guzman}, 99133E, \dodoi{10.1117/12.2233418}

\bibitem[{{Johns-Krull}(2007)}]{Johns-Krull_BFieldsTTauStars_2007ApJ}
{Johns-Krull}, C.~M. 2007, \apj, 664, 975, \dodoi{10.1086/519017}

\bibitem[{{Johnson} {et~al.}(2022){Johnson}, {David}, {Petigura}, {Isaacson},
  {Van Zandt}, {Ilyin}, {Strassmeier}, {Mallonn}, {Zhou}, {Mann}, {Livingston},
  {Luger}, {Dai}, {Weiss}, {Mo{\v{c}}nik}, {Giacalone}, {Hill}, {Rice},
  {Blunt}, {Rubenzahl}, {Dalba}, {Esquerdo}, {Berlind}, {Calkins}, \&
  {Foreman-Mackey}}]{Johnson:2022aa}
{Johnson}, M.~C., {David}, T.~J., {Petigura}, E.~A., {et~al.} 2022, \aj, 163,
  247, \dodoi{10.3847/1538-3881/ac6271}

\bibitem[{{Klein} {et~al.}(2022){Klein}, {Zicher}, {Kavanagh}, {Nielsen},
  {Aigrain}, {Vidotto}, {Barrag{\'a}n}, {Strugarek}, {Nicholson}, {Donati}, \&
  {Bouvier}}]{Klein:2022aa}
{Klein}, B., {Zicher}, N., {Kavanagh}, R.~D., {et~al.} 2022, \mnras, 512, 5067,
  \dodoi{10.1093/mnras/stac761}

\bibitem[{{Lanza} {et~al.}(1994){Lanza}, {Rodono}, \& {Zappala}}]{Lanza1994aa}
{Lanza}, A.~F., {Rodono}, M., \& {Zappala}, R.~A. 1994, \aap, 290, 861

\bibitem[{{Lightkurve Collaboration} {et~al.}(2018){Lightkurve Collaboration},
  {Cardoso}, {Hedges}, {Gully-Santiago}, {Saunders}, {Cody}, {Barclay}, {Hall},
  {Sagear}, {Turtelboom}, {Zhang}, {Tzanidakis}, {Mighell}, {Coughlin}, {Bell},
  {Berta-Thompson}, {Williams}, {Dotson}, \& {Barentsen}}]{lightkurve:2018aa}
{Lightkurve Collaboration}, {Cardoso}, J.~V.~d.~M., {Hedges}, C., {et~al.}
  2018, {Lightkurve: Kepler and TESS time series analysis in Python},
  Astrophysics Source Code Library.
\newblock \doeprint{1812.013}

\bibitem[{{Lithwick} {et~al.}(2012){Lithwick}, {Xie}, \&
  {Wu}}]{Lithwick:2012aa}
{Lithwick}, Y., {Xie}, J., \& {Wu}, Y. 2012, \apj, 761, 122,
  \dodoi{10.1088/0004-637X/761/2/122}

\bibitem[{{Lopez} {et~al.}(2012){Lopez}, {Fortney}, \& {Miller}}]{Lopez:2012aa}
{Lopez}, E.~D., {Fortney}, J.~J., \& {Miller}, N. 2012, \apj, 761, 59,
  \dodoi{10.1088/0004-637X/761/1/59}

\bibitem[{{L{\'o}pez-Morales} {et~al.}(2016){L{\'o}pez-Morales}, {Haywood},
  {Coughlin}, {Zeng}, {Buchhave}, {Giles}, {Affer}, {Bonomo}, {Charbonneau},
  {Collier Cameron}, {Consentino}, {Dressing}, {Dumusque}, {Figueira},
  {Fiorenzano}, {Harutyunyan}, {Johnson}, {Latham}, {Lopez}, {Lovis},
  {Malavolta}, {Mayor}, {Micela}, {Molinari}, {Mortier}, {Motalebi},
  {Nascimbeni}, {Pepe}, {Phillips}, {Piotto}, {Pollacco}, {Queloz}, {Rice},
  {Sasselov}, {Segransan}, {Sozzetti}, {Udry}, {Vanderburg}, \&
  {Watson}}]{Morales:2016aa}
{L{\'o}pez-Morales}, M., {Haywood}, R.~D., {Coughlin}, J.~L., {et~al.} 2016,
  \aj, 152, 204, \dodoi{10.3847/0004-6256/152/6/204}

\bibitem[{{Luger} {et~al.}(2016){Luger}, {Agol}, {Kruse}, {Barnes}, {Becker},
  {Foreman-Mackey}, \& {Deming}}]{Luger:2016aa}
{Luger}, R., {Agol}, E., {Kruse}, E., {et~al.} 2016, \aj, 152, 100,
  \dodoi{10.3847/0004-6256/152/4/10010.48550/arXiv.1607.00524}

\bibitem[{{Luger} {et~al.}(2018){Luger}, {Kruse}, {Foreman-Mackey}, {Agol}, \&
  {Saunders}}]{Luger:2018aa}
{Luger}, R., {Kruse}, E., {Foreman-Mackey}, D., {Agol}, E., \& {Saunders}, N.
  2018, \aj, 156, 99, \dodoi{10.3847/1538-3881/aad23010.48550/arXiv.1702.05488}

\bibitem[{Mahmud {et~al.}(2011)Mahmud, Crockett, Johns-Krull, Prato, Hartigan,
  Jaffe, \& Beichman}]{mahmud_starspot-induced_2011}
Mahmud, N.~I., Crockett, C.~J., Johns-Krull, C.~M., {et~al.} 2011, The
  Astrophysical Journal, 736, 123, \dodoi{10.1088/0004-637X/736/2/123}

\bibitem[{{Mann} {et~al.}(2022){Mann}, {Wood}, {Schmidt}, {Barber}, {Owen},
  {Tofflemire}, {Newton}, {Mamajek}, {Bush}, {Mace}, {Kraus}, {Thao},
  {Vanderburg}, {Llama}, {Johns-Krull}, {Prato}, {Stahl}, {Tang}, {Fields},
  {Collins}, {Collins}, {Gan}, {Jensen}, {Kamler}, {Schwarz}, {Furlan},
  {Gnilka}, {Howell}, {Lester}, {Owens}, {Suarez}, {Mekarnia}, {Guillot},
  {Abe}, {Triaud}, {Johnson}, {Milburn}, {Rizzuto}, {Quinn}, {Kerr}, {Ricker},
  {Vanderspek}, {Latham}, {Seager}, {Winn}, {Jenkins}, {Guerrero}, {Shporer},
  {Schlieder}, {McLean}, \& {Wohler}}]{Mann:2022aa}
{Mann}, A.~W., {Wood}, M.~L., {Schmidt}, S.~P., {et~al.} 2022, \aj, 163, 156,
  \dodoi{10.3847/1538-3881/ac511d}

\bibitem[{{Marley} {et~al.}(2007){Marley}, {Fortney}, {Hubickyj},
  {Bodenheimer}, \& {Lissauer}}]{Marley:2007aa}
{Marley}, M.~S., {Fortney}, J.~J., {Hubickyj}, O., {Bodenheimer}, P., \&
  {Lissauer}, J.~J. 2007, \apj, 655, 541, \dodoi{10.1086/509759}

\bibitem[{{Namekata} {et~al.}(2019){Namekata}, {Maehara}, {Notsu}, {Toriumi},
  {Hayakawa}, {Ikuta}, {Notsu}, {Honda}, {Nogami}, \&
  {Shibata}}]{Namekata_DecayRatesOfStarspots_2019ApJ}
{Namekata}, K., {Maehara}, H., {Notsu}, Y., {et~al.} 2019, \apj, 871, 187,
  \dodoi{10.3847/1538-4357/aaf471}

\bibitem[{{Nava} {et~al.}(2020){Nava}, {L{\'o}pez-Morales}, {Haywood}, \&
  {Giles}}]{Nava:2020aa}
{Nava}, C., {L{\'o}pez-Morales}, M., {Haywood}, R.~D., \& {Giles}, H. A.~C.
  2020, \aj, 159, 23, \dodoi{10.3847/1538-3881/ab53ec}

\bibitem[{{Newton} {et~al.}(2019){Newton}, {Mann}, {Tofflemire}, {Pearce},
  {Rizzuto}, {Vanderburg}, {Martinez}, {Wang}, {Ruffio}, {Kraus}, {Johnson},
  {Thao}, {Wood}, {Rampalli}, {Nielsen}, {Collins}, {Dragomir}, {Hellier},
  {Anderson}, {Barclay}, {Brown}, {Feiden}, {Hart}, {Isopi}, {Kielkopf},
  {Mallia}, {Nelson}, {Rodriguez}, {Stockdale}, {Waite}, {Wright}, {Lissauer},
  {Ricker}, {Vanderspek}, {Latham}, {Seager}, {Winn}, {Jenkins}, {Bouma},
  {Burke}, {Davies}, {Fausnaugh}, {Li}, {Morris}, {Mukai}, {Villase{\~n}or},
  {Villeneuva}, {De Rosa}, {Macintosh}, {Mengel}, {Okumura}, \&
  {Wittenmyer}}]{Newton:2019aa}
{Newton}, E.~R., {Mann}, A.~W., {Tofflemire}, B.~M., {et~al.} 2019, \apjl, 880,
  L17, \dodoi{10.3847/2041-8213/ab2988}

\bibitem[{{Owen}(2020)}]{Owen:2020aa}
{Owen}, J.~E. 2020, \mnras, 498, 5030, \dodoi{10.1093/mnras/staa2784}

\bibitem[{{Owen} \& {Wu}(2013)}]{Owen:2013aa}
{Owen}, J.~E., \& {Wu}, Y. 2013, \apj, 775, 105,
  \dodoi{10.1088/0004-637X/775/2/105}

\bibitem[{{Pepe} {et~al.}(2013){Pepe}, {Cameron}, {Latham}, {Molinari}, {Udry},
  {Bonomo}, {Buchhave}, {Charbonneau}, {Cosentino}, {Dressing}, {Dumusque},
  {Figueira}, {Fiorenzano}, {Gettel}, {Harutyunyan}, {Haywood}, {Horne},
  {Lopez-Morales}, {Lovis}, {Malavolta}, {Mayor}, {Micela}, {Motalebi},
  {Nascimbeni}, {Phillips}, {Piotto}, {Pollacco}, {Queloz}, {Rice}, {Sasselov},
  {S{\'e}gransan}, {Sozzetti}, {Szentgyorgyi}, \& {Watson}}]{Pepe:2013aa}
{Pepe}, F., {Cameron}, A.~C., {Latham}, D.~W., {et~al.} 2013, \nat, 503, 377,
  \dodoi{10.1038/nature12768}

\bibitem[{{Petigura} {et~al.}(2020){Petigura}, {Livingston}, {Batygin},
  {Mills}, {Werner}, {Isaacson}, {Fulton}, {Howard}, {Weiss}, {Espinoza},
  {Jontof-Hutter}, {Shporer}, {Bayliss}, \& {Barros}}]{Petigura:2020aa}
{Petigura}, E.~A., {Livingston}, J., {Batygin}, K., {et~al.} 2020, \aj, 159, 2,
  \dodoi{10.3847/1538-3881/ab5220}

\bibitem[{{Plavchan} {et~al.}(2020){Plavchan}, {Barclay}, {Gagn{\'e}}, {Gao},
  {Cale}, {Matzko}, {Dragomir}, {Quinn}, {Feliz}, {Stassun}, {Crossfield},
  {Berardo}, {Latham}, {Tieu}, {Anglada-Escud{\'e}}, {Ricker}, {Vanderspek},
  {Seager}, {Winn}, {Jenkins}, {Rinehart}, {Krishnamurthy}, {Dynes}, {Doty},
  {Adams}, {Afanasev}, {Beichman}, {Bottom}, {Bowler}, {Brinkworth}, {Brown},
  {Cancino}, {Ciardi}, {Clampin}, {Clark}, {Collins}, {Davison},
  {Foreman-Mackey}, {Furlan}, {Gaidos}, {Geneser}, {Giddens}, {Gilbert},
  {Hall}, {Hellier}, {Henry}, {Horner}, {Howard}, {Huang}, {Huber}, {Kane},
  {Kenworthy}, {Kielkopf}, {Kipping}, {Klenke}, {Kruse}, {Latouf}, {Lowrance},
  {Mennesson}, {Mengel}, {Mills}, {Morton}, {Narita}, {Newton}, {Nishimoto},
  {Okumura}, {Palle}, {Pepper}, {Quintana}, {Roberge}, {Roccatagliata},
  {Schlieder}, {Tanner}, {Teske}, {Tinney}, {Vanderburg}, {von Braun}, {Walp},
  {Wang}, {Wang}, {Weigand}, {White}, {Wittenmyer}, {Wright}, {Youngblood},
  {Zhang}, \& {Zilberman}}]{Plavchan2020aa}
{Plavchan}, P., {Barclay}, T., {Gagn{\'e}}, J., {et~al.} 2020, \nat, 582, 497,
  \dodoi{10.1038/s41586-020-2400-z}

\bibitem[{{Prato} {et~al.}(2008){Prato}, {Huerta}, {Johns-Krull}, {Mahmud},
  {Jaffe}, \& {Hartigan}}]{Prato_VisIR_2008ApJ}
{Prato}, L., {Huerta}, M., {Johns-Krull}, C.~M., {et~al.} 2008, \apjl, 687,
  L103, \dodoi{10.1086/59320110.48550/arXiv.0809.3599}

\bibitem[{{Rajpaul} {et~al.}(2015){Rajpaul}, {Aigrain}, {Osborne}, {Reece}, \&
  {Roberts}}]{Rajpaul:2015aa}
{Rajpaul}, V., {Aigrain}, S., {Osborne}, M.~A., {Reece}, S., \& {Roberts}, S.
  2015, \mnras, 452, 2269, \dodoi{10.1093/mnras/stv1428}

\bibitem[{Rasmussen \& Williams(2006)}]{Rasmussen:2006aa}
Rasmussen, C.~E., \& Williams, C. K.~I. 2006, Gaussian processes for machine
  learning., Adaptive computation and machine learning (MIT Press), I--XVIII,
  1--248

\bibitem[{{Rizzuto} {et~al.}(2020){Rizzuto}, {Newton}, {Mann}, {Tofflemire},
  {Vanderburg}, {Kraus}, {Wood}, {Quinn}, {Zhou}, {Thao}, {Law}, {Ziegler}, \&
  {Brice{\~n}o}}]{Rizzuto:2020aa}
{Rizzuto}, A.~C., {Newton}, E.~R., {Mann}, A.~W., {et~al.} 2020, \aj, 160, 33,
  \dodoi{10.3847/1538-3881/ab94b7}

\bibitem[{{Saar} \& {Donahue}(1997)}]{Saar_RVVariation_1997ApJ}
{Saar}, S.~H., \& {Donahue}, R.~A. 1997, \apj, 485, 319, \dodoi{10.1086/304392}

\bibitem[{{Su{\'a}rez Mascare{\~n}o} {et~al.}(2021){Su{\'a}rez Mascare{\~n}o},
  {Damasso}, {Lodieu}, {Sozzetti}, {B{\'e}jar}, {Benatti}, {Zapatero Osorio},
  {Micela}, {Rebolo}, {Desidera}, {Murgas}, {Claudi}, {Gonz{\'a}lez
  Hern{\'a}ndez}, {Malavolta}, {del Burgo}, {D'Orazi}, {Amado}, {Locci},
  {Tabernero}, {Marzari}, {Aguado}, {Turrini}, {Cardona Guill{\'e}n},
  {Toledo-Padr{\'o}n}, {Maggio}, {Aceituno}, {Bauer}, {Caballero},
  {Chinchilla}, {Esparza-Borges}, {Gonz{\'a}lez-{\'A}lvarez}, {Granzer},
  {Luque}, {Mart{\'\i}n}, {Nowak}, {Oshagh}, {Pall{\'e}}, {Parviainen},
  {Quirrenbach}, {Reiners}, {Ribas}, {Strassmeier}, {Weber}, \&
  {Mallonn}}]{SM21}
{Su{\'a}rez Mascare{\~n}o}, A., {Damasso}, M., {Lodieu}, N., {et~al.} 2021,
  Nature Astronomy, 6, 232, \dodoi{10.1038/s41550-021-01533-7}

\bibitem[{{Tejada Arevalo} {et~al.}(2022){Tejada Arevalo}, {Tamayo}, \&
  {Cranmer}}]{Tejada:2022aa}
{Tejada Arevalo}, R., {Tamayo}, D., \& {Cranmer}, M. 2022, \apjl, 932, L12,
  \dodoi{10.3847/2041-8213/ac70e0}

\bibitem[{Virtanen {et~al.}(2020)Virtanen, Gommers, Oliphant, Haberland, Reddy,
  Cournapeau, Burovski, Peterson, Weckesser, Bright, {van der Walt}, Brett,
  Wilson, Millman, Mayorov, Nelson, Jones, Kern, Larson, Carey, Polat, Feng,
  Moore, {VanderPlas}, Laxalde, Perktold, Cimrman, Henriksen, Quintero, Harris,
  Archibald, Ribeiro, Pedregosa, {van Mulbregt}, \& {SciPy 1.0
  Contributors}}]{scipy}
Virtanen, P., Gommers, R., Oliphant, T.~E., {et~al.} 2020, Nature Methods, 17,
  261, \dodoi{10.1038/s41592-019-0686-2}

\bibitem[{{Vogt} {et~al.}(1994){Vogt}, {Allen}, {Bigelow}, {Bresee}, {Brown},
  {Cantrall}, {Conrad}, {Couture}, {Delaney}, {Epps}, {Hilyard}, {Hilyard},
  {Horn}, {Jern}, {Kanto}, {Keane}, {Kibrick}, {Lewis}, {Osborne},
  {Pardeilhan}, {Pfister}, {Ricketts}, {Robinson}, {Stover}, {Tucker}, {Ward},
  \& {Wei}}]{Vogt:1994aa}
{Vogt}, S.~S., {Allen}, S.~L., {Bigelow}, B.~C., {et~al.} 1994, in Society of
  Photo-Optical Instrumentation Engineers (SPIE) Conference Series, Vol. 2198,
  \procspie, ed. D.~L. {Crawford} \& E.~R. {Craine}, 362,
  \dodoi{10.1117/12.176725}

\bibitem[{{W}es {M}c{K}inney(2010)}]{pandas}
{W}es {M}c{K}inney. 2010, in {P}roceedings of the 9th {P}ython in {S}cience
  {C}onference, ed. {S}t\'efan van~der {W}alt \& {J}arrod {M}illman, 56 -- 61,
  \dodoi{10.25080/Majora-92bf1922-00a}

\bibitem[{{Yu} {et~al.}(2019{\natexlab{a}}){Yu}, {Donati}, {Grankin}, {Collier
  Cameron}, {Moutou}, {Hussain}, {Baruteau}, {Jouve}, \& {MaTYSSE
  Collaboration}}]{Yu_V410tau_2019}
{Yu}, L., {Donati}, J.~F., {Grankin}, K., {et~al.} 2019{\natexlab{a}}, \mnras,
  489, 5556, \dodoi{10.1093/mnras/stz2481}

\bibitem[{{Yu} {et~al.}(2019{\natexlab{b}}){Yu}, {Donati}, {Grankin}, {Collier
  Cameron}, {Moutou}, {Hussain}, {Baruteau}, {Jouve}, \& {MaTYSSE
  Collaboration}}]{Yu_v410Tau_2019MNRAS}
---. 2019{\natexlab{b}}, \mnras, 489, 5556, \dodoi{10.1093/mnras/stz2481}

\bibitem[{{Zicher} {et~al.}(2022){Zicher}, {Barrag{\'a}n}, {Klein}, {Aigrain},
  {Owen}, {Gandolfi}, {Lagrange}, {Serrano}, {Kaye}, {Nielsen}, {Rajpaul},
  {Grandjean}, {Goffo}, \& {Nicholson}}]{Zicher:2022aa}
{Zicher}, N., {Barrag{\'a}n}, O., {Klein}, B., {et~al.} 2022, \mnras, 512,
  3060, \dodoi{10.1093/mnras/stac614}

\end{thebibliography}

\end{document}